\documentclass[aps,prx,twocolumn,superscriptaddress,longbibliography]{revtex4-1}
\usepackage{graphicx}
\usepackage{latexsym}
\usepackage{amssymb}
\usepackage{amsmath}
\usepackage{amsfonts}
\usepackage{upgreek}
\usepackage{bm}
\usepackage{multirow}
\usepackage{color}
\usepackage{hyperref}
\hypersetup{
colorlinks = true,
linkcolor = [rgb]{0.70,0.13,0.13},
citecolor = [rgb]{0.13,0.55,0.13},
urlcolor = [rgb]{0.25, 0.41, 0.88}}

\newcommand{\dd}{\mathrm{d}}
\newcommand{\ii}{\mathrm{i}}

\newcommand{\U}{\mathrm{U}}
\newcommand{\SU}{\mathrm{SU}}
\renewcommand{\O}{\mathrm{O}}
\newcommand{\SO}{\mathrm{SO}}

\newcommand{\dsZ}{\mathbb{Z}}

\newcommand{\scO}{\mathcal{O}}
\newcommand{\scL}{\mathcal{L}}
\newcommand{\scH}{\mathcal{H}}
\newcommand{\scK}{\mathcal{K}}
\newcommand{\scP}{\mathcal{P}}
\newcommand{\scT}{\mathcal{T}}

\renewcommand{\Re}{\operatorname{Re}}
\renewcommand{\Im}{\operatorname{Im}}
\newcommand{\vect}[1]{{\bm{#1}}}

\newcommand{\mat}[1]{\left[\begin{matrix}#1\end{matrix}\right]}
\newcommand{\smat}[1]{\left[\begin{smallmatrix}#1\end{smallmatrix}\right]}
\newcommand{\cmat}[1]{\left\{\begin{matrix}#1\end{matrix}\right\}}
\newcommand{\eq}[1]{\begin{equation}#1\end{equation}}
\newcommand{\eqs}[1]{\begin{equation}\begin{split}#1\end{split}\end{equation}}
\newcommand{\eqnref}[1]{Eq.\,\eqref{#1}}
\newcommand{\figref}[1]{Fig.\,\ref{#1}}
\newcommand{\tabref}[1]{Tab.\,\ref{#1}}

\begin{document}
\title{Emergent Symmetry and Conserved Current at a One Dimensional Incarnation of Deconfined Quantum Critical Point}

\author{Rui-Zhen Huang}
\affiliation{Kavli Institute for Theoretical Sciences, University of Chinese Academy of Sciences, Beijing 100190, China}
\author{Da-Chuan Lu}
\affiliation{Department of Physics, University of California at San Diego, La Jolla, CA 92093, USA}
\author{Yi-Zhuang You}
\affiliation{Department of Physics, University of California at San Diego, La Jolla, CA 92093, USA}
\author{Zi Yang Meng}
\affiliation{Beijing National Laboratory for Condensed Matter Physics and Institute of Physics, Chinese Academy of Sciences, Beijing 100190, China}
\affiliation{Songshan Lake Materials Laboratory, Dongguan, Guangdong 523808, China}
\affiliation{Department of Physics, The University of Hong Kong, China}
\author{Tao Xiang}
\affiliation{Beijing National Laboratory for Condensed Matter Physics and Institute of Physics, Chinese Academy of Sciences, Beijing 100190, China}
\affiliation{Collaborative Innovation Center of Quantum Matter, Beijing 100190, China}
\date{\today}

\begin{abstract}
The deconfined quantum critical point (DQCP) was originally proposed as a continuous transition between two spontaneous symmetry breaking phases in 2D spin-1/2 systems. While great efforts have been spent on the DQCP for 2D systems, both theoretically and numerically, ambiguities among the nature of the transition are still not completely clarified. Here we shift the focus to a recently proposed 1D incarnation of DQCP in a spin-1/2 chain. By solving it with the variational matrix product state in the thermodynamic limit, a continuous transition between a valence-bond solid phase and a ferromagnetic phase is discovered. The scaling dimensions of various operators are calculated and compared with those from field theoretical description. At the critical point, two emergent $\O(2)$ symmetries are revealed, and the associated conserved current operators with exact integer scaling dimensions are determined with scrutiny. Our findings provide the low-dimensional analog of DQCP where unbiased numerical results are in perfect agreement with the controlled field theoretical predictions and have extended the realm of the unconventional phase transition as well as its identification with the advanced numerical methodology.
\end{abstract}

\maketitle

\section{\label{intro}Introduction}

The deconfined quantum critical point (DQCP)\cite{ashvinlesik,deconfine1,deconfine2} was originally proposed as a continuous quantum phase transition between two spontaneous symmetry breaking (SSB) phases, such as the N\'eel antiferromagnetic (AFM) phase and the valance bond solid (VBS) phase in (2+1)D quantum magnets. In the conventional Landau-Ginzburg-Wilson (LGW) paradigm, such scenario can not occur without fine tuning. Since the AFM and the VBS ordering transitions are independent as they breaks very different symmetries, the transitions would generally happen at two separate points, leaving an intermediate disordered or coexistent phase. However, various theoretical and numerical studies\cite{Sandvik2007,Melko2008,Charrier2008,Kuklov08,Chen2009,Lou2009,Charrier2010,Banerjee2010,Sandvik2010,Nahum2011,Bartosch2013,Harada2013,Chen2013,Block2013,Sreejith2015,Nahum2015a,Nahum2015b,Shao2016,Shao2017b,YQQin2017,Sato2017,NvsenMa2018,BowenZhao2018,Serna2018,Ippoliti2018} show that there could be a continuous (or weakly first order) direct transition between the two phases. The occurrence of DQCP without fine tuning is consistent with the constraint imposed by the Lieb-Schultz-Mattis (LSM) theorem\cite{LSM} and its generalizations\cite{Oshikawa2000,Hastings2004,Watanabe2015,Cheng2016,Po2017,Qi2017,Lu2017a,Lu2017b,Yang2018,Cheng2018}. The (generalized) LSM theorem asserts that a translationally invariant spin system with spin-1/2 (projective representation of on-site symmetry) per unit cell can not admit a featureless (fully-gapped and non-degenerated) symmetric ground state, which rules out the conventional phase transition between an SSB phase and a featureless symmetric phase in such systems, paving way for the DQCP, although the LSM theorem does not rule out other possibilities like a co-existence phase, a first-order transition, or a topological ordered spin liquid. Recent theoretical developments\cite{Cho2017,Komargodski2017,YCWang2017,Komargodski2018,Metlitski2018,Jian2018,YCWang2017} further relates the DQCP in LSM systems to the boundary of symmetry protected topological states in one-higher-dimension, where the LSM theorem is manifested as the symmetry anomaly in the field theory description when the lattice symmetry is encoded as part of the internal symmetry. The anomaly matching requires that the symmetric state of such quantum systems must either be critical (as the DQCP) or possess topological order.

\begin{figure}[tbp]
\includegraphics[width=0.8\columnwidth]{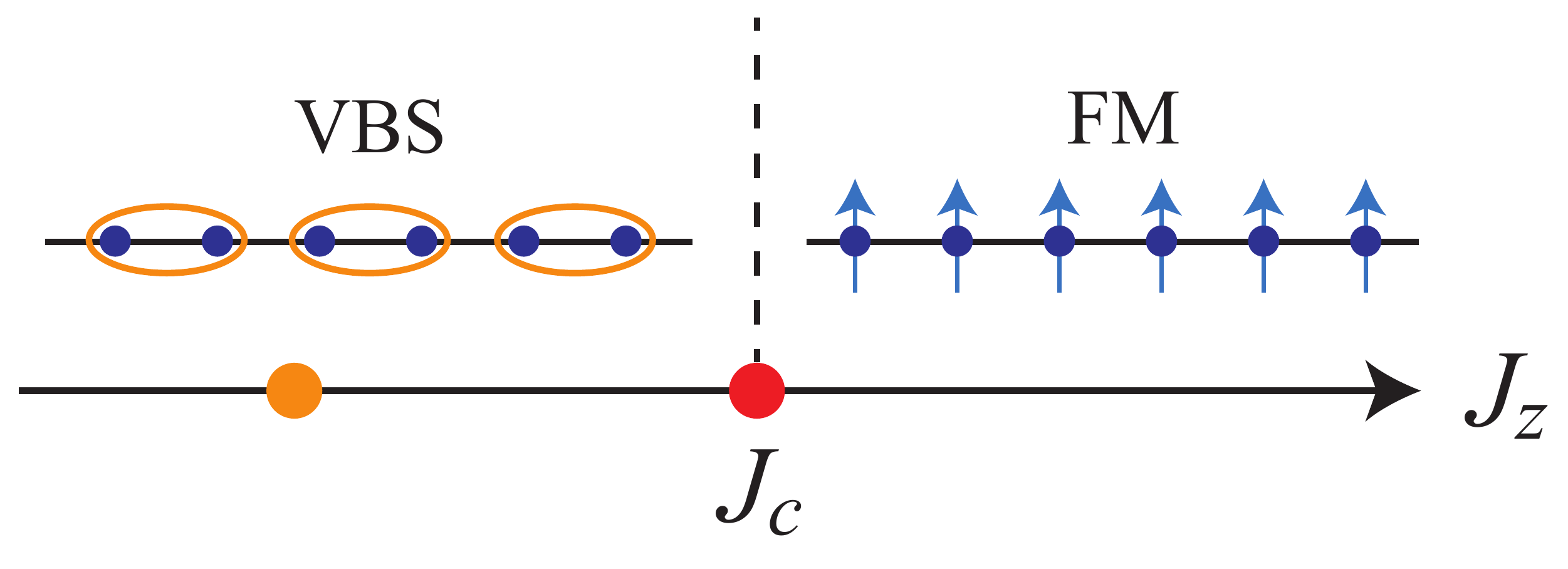}
\caption{Schematic phase diagram for the model in \eqnref{eq:model}. When $J_z < J_c$ the system is inside a VBS phase with translational symmetry breaking and when $J_z>J_c$ the system is inside a FM phase with $\dsZ_2^x$ on-site symmetry breaking. At $J_c$, an incarnation of DQCP emerges. In the VBS phase there exists an exactly solvable point at $J_z=1$, where the ground state is a product state of dimerised neighbor spins(orange point).}
\label{fig:phasediagram}
\end{figure}

In this work we explore an (1+1)D analog of DQCP in a quantum spin chain. We pick up the model proposed by Jiang and Motrunich \cite{SHJiang2019}, which is a spin chain with spin-1/2 per site, described by the following Hamiltonian
\begin{equation}
\begin{split}
H = \sum_i&(-J_x S^x_i S^x_{i+1} -J_z S^z_i S^z_{i+1}) \\&+ (K_x S^x_i S^x_{i+2} + K_z S^z_i S^z_{i+2}).
\end{split}
\label{eq:model}
\end{equation}
The model respects the lattice translation symmetry and the on-site $\dsZ_2^x\times\dsZ_2^z$ spin flip symmetry, where $\dsZ_2^x$ ($\dsZ_2^z$) corresponds to the two-fold spin rotation around the $x$-axis ($z$-axis), which are subgroups of the full $\SO(3)$ spin rotation symmetry. The spin on each site forms a projective representation of $\dsZ_2^x\times\dsZ_2^z$, hence the LSM theorem is in operation to forbid the gapped symmetric ground state for any choice of the model parameters. For the sake of simplicity, we fix the second neighbor antiferromagnetic interaction $K_x=K_z=1/2$ and the nearest neighbor ferromagnetic interaction $J_x=1$, therefore, the only driving parameter is the nearest neighbor ferromagnetic $J_z$. In Ref.~\cite{SHJiang2019}, it is argued that there is a continuous quantum phase transition between a VBS phase ($J_z \sim 1$) and a spin-$z$ ordered ferromagnetic ($z$-FM) phase ($J_z \gg 1$), where the VBS  breaks the translational symmetry and $z$-FM  breaks the $\dsZ_2^x$ on-site symmetry. The transition is analogous to the 2D DQCP in the sense that it is also a direct continuous transition between two different SSB phases is the LSM system\cite{SHJiang2019}. If the transition preserves the full symmetry, it must be quantum critical by the LSM theorem. This 1D example provides us opportunities to uncover common features of DQCP, such as the emergent symmetries and the associated conserved current fluctuations, that are shared between different dimensions.

We also note that the model proposed in Ref.~\cite{SHJiang2019} is not the only example of 1D analog of DQCP. Emergent continuous symmetry in quantum spin chain systems have already been discussed in Refs.~\cite{AffleckHaldane1987} and revisited numerically and analytically in Refs.~\cite{Patil2018,Mudry2019}, and in Refs.~\cite{Sengupta2002,Sandvik2004}, a 1D extended Hubbard model were studied with quantum Monte Carlo and a direct continuous transition between a charge density wave (CDW) phase and a bond order wave (BOW) phase is found. The CDW and BOW phases respectively break the bond-centered and site-centered reflection symmetries. The lattice symmetries together with the charge $\U(1)$ symmetry (at a fractional filling) also lead to a generalized LSM constraint~\cite{Cho2017} that ensures the transition to be critical. So the CDW-BOW transition also qualifies as a 1D analog of DQCP. However, as will be explained later in the text, the CDW and BOW transition in Refs.~\cite{Sengupta2002,Sandvik2004} is  not discussed in the language of DQCP and the unique features such as emergent continuous symmetry and conserved current at the critical point, were not investigated.

Here we employed variational matrix product state to solve the spin model in \eqnref{eq:model} at the thermodynamic limit. The ground state phase diagram is determined unambiguously. As depicted in \figref{fig:phasediagram}, there exists a direct continuous transition between the VBS and the $z$-FM phases. The critical point locates at $J_c=1.4645$ for our choice of parameters $J_x=2K_x=2K_z=1$. Since the critical point is described by a Luttinger liquid with continuously tunable Luttinger parameters (realized by tuning $J_x,K_x,K_z$), the critical exponents are not universal. However, all critical exponents are controlled by a single Luttinger parameter, which posts non-trivial consistency relations among the exponents. We calculated the scaling dimensions of various operators at this critical point. We can verify that they all point to a consistent result of the Luttinger parameter, in agreement with the field theory expectation.

Furthermore, we found two emergent continuous $\O(2)$ symmetries at the critical point. They separately correspond to the rotation between VBS and $z$-FM fluctuations and the rotation between $x$-FM and $y$-AFM fluctuations. The evidences are two-folded: (i) the scaling dimensions for the critical fluctuations related by the emergent symmetry are identical, and (ii) the associated emergent conserved currents dictated by the Noether theorem are observed with their scaling dimensions precisely pinned at integer value. The emergent continuous symmetry is a robust feature of the DQCP~\cite{Sandvik2007,Nahum2015b,YQQin2017,Sato2017,NvsenMa2018Current}. For the 2D case, there are accumulating evidence that the emergent symmetry can manifest itself even if the transition is weakly first order \cite{BowenZhao2018,Wildeboer2018,Serna2018,NvsenMa2018Current}. It is further proposed that the associated emergent conserved current fluctuations could serve as a hallmark of the DQCP in candidate materials.\cite{Lee2019} To test these ideas in 1D, we identify the microscopic representations of current operators, and measure their scaling dimensions in numerics. We found that the current operators have exact scaling dimensions $\Delta=1$, indicating the conservation of these currents, which provides compelling evidence for the emergent $\O(2)\times\O(2)$ symmetry at our critical point.

The rest of this paper is organized as follows. In Sec.~\ref{sec:fieldtheory}, the field theoretical description of the 1D DQCP is presented, with the scaling behavior of relevent fields (the order parameter, conserved current operator) discussed in details. Different from Ref.~\cite{SHJiang2019}, we adopt to a formulation of non-linear $\sigma$-model (NLSM) with a Wess-Zumino-Witten (WZW) term. The numerical method to study the ground state and critical properties is introduced in Sec.~\ref{sec:Method}. In Sec.~\ref{sec:FLS} we show the critical properties at the DQCP by systematic finite length scaling analysis with the critical exponents determined at high precision. Then in Sec.~\ref{sec:Symmetry}, we study the emergent $\O(2)\times\O(2)$ symmetry and the associated conserved currents correlation at the DQCP. Finally a summary is given in Sec.~\ref{sec:Summary}.

\section{Field theoretical description}
\label{sec:fieldtheory}

Many field theory descriptions
\cite{ashvinlesik,deconfine1,deconfine2,hermele2005,ranwen,senthilfisher,groverashvin,lulee,xudual,karchtong,seiberg2,mengxu,SO5,Assaad2016,maxryan17,youSMG2}
have been proposed for the 2D DQCP which are believed to be equivalent (or dual) to each other at low energy, including the non-linear $\sigma$-model (NLSM)\cite{senthilfisher}, the non-compact CP$^1$ (NCCP$^1$)
theory\cite{deconfine1,deconfine2} and various versions of the quantum electrodynamics (QED) or quantum chromodynamics (QCD) theories \cite{SO5,Assaad2016,maxryan17,youSMG2,Xu2018}. In parallel to the situation of 2D DQCP, there are also low-dimensional correspondences of all these descriptions for 1D DQCP. Much of them have been discussed in Ref.~\cite{SHJiang2019} in great details. However, here we point out a field theory description in terms of an anisotropic $\O(4)$ NLSM in (1+1)D with a WZW term at level $k=1$, which has not been much analyzed in Ref.~\cite{SHJiang2019}. This description (and its fermionic spinon construction) provides us a particularly convenient approach to identify the emergent symmetry and conserved currents at the 1D DQCP. The Lagrangian in the Euclidean space-time reads
\eqs{\label{eq:NLSM}
\scL[\vect{n}]=&\frac{1}{2\kappa}(\partial_\mu\vect{n})^2
+\frac{\ii k}{2\pi^2} \epsilon^{abcd}n_a\partial_\tau n_b\partial_x n_c\partial_u n_d\\
+&\lambda (n_3^2-n_4^2)+\lambda' (n_1^2-n_2^2)\\
+&\mu(n_1^2+n_2^2-n_3^2-n_4^2)+\cdots,
}
where $\vect{n}=(n_1,n_2,n_3,n_4)$ is a four-component real unit vector defined at each space-time position $x^\mu=(\tau,x)$\footnote{In the WZW term, the $\vect{n}$ is further suspended to an extra dimension $u$, such that $\vect{n}(\tau,x,u=0)\equiv(0,0,0,1)$ is smoothly continued to $\vect{n}(\tau,x,u=1)=\vect{n}(\tau,x)$.}. The components of this $\O(4)$ vector $\vect{n}$ parameterize the four leading ordering tendencies ($x$-FM, $y$-AFM, $z$-FM, VBS) in the spin model. They can be represented by the microscopic spin operators as
\eqs{\label{eq: n=S}
\text{$x$-FM: }&n_1\sim S_i^x,\\
\text{$y$-AFM: }&n_2\sim(-)^i S_i^y,\\
\text{$z$-FM: }&n_3\sim S_i^z,\\
\text{VBS: }&n_4\sim(-)^i\vect{S}_i\cdot\vect{S}_{i+1}.}
None of them gets ordered at the DQCP, but they exhibit strong critical fluctuations at low-energy. In \eqnref{eq:NLSM}, the first term $(\partial_\mu\vect{n})^2=(\partial_\tau \vect{n})^2+(\partial_x \vect{n})^2$ describes the kinetic energy of their critical fluctuation. The WZW term at level $k=1$ introduces a symmetry anomaly which is crucial to capture the obstruction to featureless symmetric state in the LSM setting. Finally, the anisotropy terms $\lambda$, $\lambda'$ and $\mu$ are introduced to break the $\O(4)$ symmetry to the microscopic $\dsZ_2^x\times\dsZ_2^z$ symmetry of the lattice spin model.

To describe the $z$-FM--VBS DQCP, one focuses on the parameter regime where $\mu>0$ is positive and $\lambda,\lambda'$ are perturbative. In particular, $\lambda$ will be the tuning parameter that drives the $z$-FM--VBS transition. The reasons are as follows. First of all, a positive $\mu$ term favors the ordering of $n_3,n_4$ ($z$-FM and VBS) over $n_1,n_2$ ($x$-FM and $y$-AFM), which places the system in adjacent to the $z$-FM and VBS ordered phases. However with the $\mu$ term only, neither the $z$-FM nor the VBS long-range order can develop, because the field theory still admits an $\O(2)_{\theta}$ symmetry that rotates $(n_3,n_4)$ as an $\O(2)$ vector, which is a continuous symmetry that can not be spontaneously broken in (1+1)D due to the Mermin-Wagner theorem. The anisotropy $\lambda$ is introduced to explicitly break the $\O(2)_{\theta}$ symmetry, such that the $(n_3,n_4)$ ordering becomes possible. When $\lambda<0$ (or $\lambda>0$), the $n_3$ (or $n_4$) ordering is favored, leading to the $z$-FM (or the VBS) phase. Therefore the transition happens at the $\lambda=0$ point. Finally, the $\lambda'$ anisotropy is generally allowed in the field theory to explicitly break the $\O(2)_\phi$ symmetry that rotates $(n_1,n_2)$ as an $\O(2)$ vector, because the microscopic spin model does not respect the $\O(2)_\phi$ symmetry indeed. However, the $\lambda'$ term is irrelevant at the DQCP when $\mu>0$, as shown in Ref.~\cite{SHJiang2019} using the Abelian bosonization technique (see also Appendix \ref{sec:bosonization}). In conclusion, \eqnref{eq:NLSM} provides a field theory description of the $z$-FM--VBS DQCP when $\lambda$ is fine-tuned to the $\lambda=0$ critical point.

The \eqnref{eq:NLSM} NLSM field theory immediately leads to a powerful prediction about the emergent symmetry at the DQCP. At the critical point, $\lambda$ is fine-tuned to zero and $\lambda'$ flows to zero under renormalization, so the $\mu$ term is the only relevant perturbation that remains in the theory, which is symmetric under $\O(2)_\phi\times \O(2)_\theta$. The symmetry groups are labeled by $\phi$ and $\theta$, because we are going to parameterize the $\O(4)$ vector $\vect{n}$ as\footnote{Here we assume the compactification condition of $\phi\sim\phi+2\pi$ and $\theta\sim\theta+2\pi$, different from Ref.\cite{SHJiang2019}.}
\eq{\label{eq: angle=n}
e^{\ii\phi}\sim n_1+\ii n_2,\quad e^{\ii\theta}\sim n_3+\ii n_4.}
Then $\O(2)_\phi$ (or $\O(2)_\theta$) corresponds to rotating or reversing the angle $\phi$ (or $\theta$), under which the $\mu$ term is clearly invariant. Therefore the $z$-FM--VBS DQCP should process an emergent $\O(2)_\phi\times \O(2)_\theta$ symmetry. As a consequence, there must be emergent conserved currents $J_\mu^\phi$ and $J_\mu^\theta$ associated to the emergent $\O(2)_\phi\times \O(2)_\theta$ symmetry,
\eq{\label{eq: J=angle}
(J_\tau^\phi,J_x^\phi)=(\ii\partial_\tau\phi,\partial_x\phi),\quad(J_\tau^\theta,J_x^\theta)=(\ii\partial_\tau\theta,\partial_x\theta).}
In terms of spin operator, the emergent conserved currents correspond to (see Appendix \ref{sec:symmetry} for derivation)
\eq{\label{eq: J=S}J_\tau^\phi\sim J_x^\theta \sim (-)^i S_i^z,\quad J_x^\phi\sim J_\tau^\theta \sim (-)^iS_i^xS_{i+1}^y.}
The components $J_\tau^\phi$ and $J_x^\theta$ both correspond to the $z$-AFM fluctuation and the components $J_x^{\phi}$ and $J_\tau^{\theta}$ can be measured as staggered modulation of the $xy$-dimmerization $S^{x}_{i}S^{y}_{i+1}$ (which will be called the $xy$-VBS fluctuation). As shown in Sec.~\ref{sec:Symmetry}, these two conserved currents are indeed present at the DQCP, in support of the emergent $\O(2)_\phi\times \O(2)_\theta$ symmetry.

Following the approach developed in Ref.~\cite{youspn}, the $\O(4)$ NLSM in \eqnref{eq:NLSM} can be Abelian bosonized to the standard Luttinger liquid theory, which allows us to calculate the scaling dimensions for all operators (see Appendix \ref{sec:bosonization} for details). The field theory calculation predicts the following scaling behavior of the correlation functions,
\eqs{
G_x(r)=\langle S_i^xS_{i+r}^x\rangle&\sim\frac{1}{r^{2/g}}+\frac{(-1)^r}{r^{2/g+g/2}},\\
G_y(r)=\langle S_i^yS_{i+r}^y\rangle&\sim\frac{1}{r^{2/g+g/2}}+\frac{(-1)^r}{r^{2/g}},\\
G_z(r)=\langle S_i^zS_{i+r}^z\rangle&\sim\frac{1}{r^{g/2}}+\frac{(-1)^r}{r^2},\\
G_\Psi(r)=\langle \Psi_{i}\Psi_{i+r}\rangle&\sim\frac{(-1)^r}{r^{g/2}},\\
G_\Gamma(r)=\langle \Gamma_{i}\Gamma_{i+r}\rangle&\sim\frac{(-1)^r}{r^2},
\label{eq:correlations}}
where $\Psi_i=\vect{S}_i\cdot\vect{S}_{i+1}$ is the ordinary dimmer operator and $\Gamma_i=S_i^xS_{i+1}^y$ is the $xy$-dimmer operator. The exponents of the correlation functions are all controlled by a single Luttinger parameter $g$, which is also related to the critical exponent $\nu$ by
\eq{\nu=\frac{1}{2-g}.
\label{eq:luttinger}}
Most notably, the exponent of the $z$-AFM fluctuation and the $xy$-VBS fluctuation are exactly pinned at the integer 2 (scaling dimension $\Delta=1$) by the emergent $\O(2)_\phi\times\O(2)_\theta$ symmetry.

In the sections followed, we will employ the variational matrix product state simulations, to verify the field theoretical predictions in \eqnref{eq: J=S}, \eqnref{eq:correlations} and \eqnref{eq:luttinger} in a step-by-step manner.

\section{Matrix product state and finite correlation length scaling}
\label{sec:Method}

We employ the matrix product state (MPS) to study the ground state properties of the model in \eqnref{eq:model}. The MPS~\cite{Schollwock2005,Cirac2006,Cirac2008,Schollwock2011} represents the quantum many-body state in a chain of $D \times D$ matrices multiplied together, whose entanglement is bounded by the bond dimension $D$. The MPS can provide very efficient representations for gapped quantum many-body states in 1D, due to the area-law scaling of entanglement.
However for gapless states at quantum critical points, the entanglement scales logarithmically, which generally requires tensor network~\cite{cite_MERA,cite_TNS_Lanczos} with more complicated structure but more powerful expressing capability to represent the state, but the computational cost for those tensor networks will also be much higher.

Within the MPS representation, (non-)abelian symmetries have been used to enlarge the effective bond dimension $D$ at the cost of a great increase of the computational complexity. Nevertheless, any finite bond dimension $D$ will still introduce a finite effective correlation length $\xi$, which inevitably prevents us to capture the critical exponents in the thermodynamic limit. Another issue in the MPS simulation is the boundary effect for a finite quantum system. Typically the entanglement entropy in the ground state is much larger for a periodic chain than a open one, therefor the open boundary condition is widely used to reduce the computation cost\cite{pbc_dmrg}. However this introduces strong boundary effects which require large system sizes to obtain the accurate ground state properties, especially for a critical one.


Well aware of the aforementioned difficulties, we take a different strategy to simulate the DQCP with MPS in this work. We use an infinite MPS trial wave function and apply the finite correlation length scaling to analyze the critical properties. In this way, the boundary effect in the finite MPS calculation is avoided and the MPS with all $D$ can be utilized for evaluating the critical exponents. Moreover, we found that correlation functions at short and long distances can be captured by the same scaling theory and hence more accurate critical exponents can be obtained by collapsing of these correlation functions.



In our calculation, we first use a variational way to optimize the MPS. In order to study the scaling behavior near the critical point, the ground state wave function must be carefully optimized for a given $D$. Moreover the long distance correlation requires higher accuracy than local physical quantities. In order to obtain such accurate MPS, we adopt the tangent space MPS technique~\cite{ref_tangent_MPS1,ref_tangent_MPS2,ref_vmps}, in which the MPS with given $D$ forms a sub-manifold in the full Hilbert space. We directly calculate the energy gradient in this sub-manifold and optimize it upon convergence to a sufficient small value. And in order to avoid local minimum, we take several independent runs with random initial states to find the best wave function. The wave function thus optimized is capable of describing not only the local physical quantities but also the long range correlations.

Then we carry out finite length scaling analysis upon the obtained MPS wave functions. To this end, one first needs to identify the finite length scale associated with the finite $D$ MPS sub-manifold, and it has been suggested recently in Refs.~\cite{cite_FLS_PEPS1,cite_FLS_PEPS2} that the correlation length $\xi$ determined by the MPS wave function at the critical point serves as this finite length scale. The finite correlation length stems from the finite entanglement properties for finite $D$ MPS, so this approach is also known as finite entanglement scaling~\cite{cite_FLS_MPS1,cite_FLS_MPS2,
cite_FLS_MPS3,cite_FLS_MPS4}. For a normalized MPS wave function with a given $D$, the effective correlation length $\xi$ can be easily evaluated as
\begin{equation}
\xi = -1/\log(\vert \lambda \vert),
\label{eq:correlationlength}
\end{equation}
where $\lambda$ is the second largest eigenvalue of the transfer matrix formed by the local matrices in the MPS~\cite{Schollwock2011}.

For a order parameter $n_a$ the finite length scaling near the critical point reads
\begin{equation}
\label{eq:Eq_scale1}
n_a(g,\xi) = \xi^{-\Delta_a} f_1(\delta \xi^{1/\nu}),
\end{equation}
where $\Delta_a$ is the scaling dimension of the order parameter, $\delta=J_z - J_c$ is the distance to the critical point and $f_1$ is the scaling function.
For a generic correlation function $G_a(r,\xi)$, at the critical point, as a function of distance $r$, its finite length scaling has the following form
\begin{equation}
\label{eq:Eq_scale2}
G_a(r,\xi) = \frac{1}{r^{\eta_a}} f_2(r/\xi),
\end{equation}
where $\eta_a=2\Delta_a$ is twice the scaling dimension of the operators involved in the correlation function and $f_2$ is another scaling function. For short range correlation $r \ll \xi$, $r/\xi$ is deflectable and $G_a(r,\xi)$ is approximated by a function of power-law form. And for long range correlation $r \gg \xi$, $G_a(r,\xi)$ is approximated by an exponential function. Here we show that at all the length scales the correlation functions satisfy the full scaling form \eqnref{eq:Eq_scale2}. Moreover by systematically calculating the ground state using different $D$ MPS, hence different $\xi$, one can perform finite correlation length scaling analysis according to \eqnref{eq:Eq_scale1} and \eqnref{eq:Eq_scale2} to obtain the critical exponents, as will be shown in details in Sec.~\ref{sec:FLS}.

We would like to emphasize that in previous tensor network studies, the finite correlation length scaling analyses have been applied on conventional continuous phase transitions~\cite{cite_FLS_PEPS1,cite_FLS_PEPS2}, namely, the corresponding critical exponents are well established. In this work we for the first time test the virtue of the finite correlation length scaling in the MPS with a non-trivial DQCP with unknown exponents, at the level of numerical methodology for MPS and tensor network, our systematic analyses establish a protocol for future research.

\section{Critical properties}
\label{sec:FLS}

\begin{figure}[htbp]
\includegraphics[width=0.9\columnwidth]{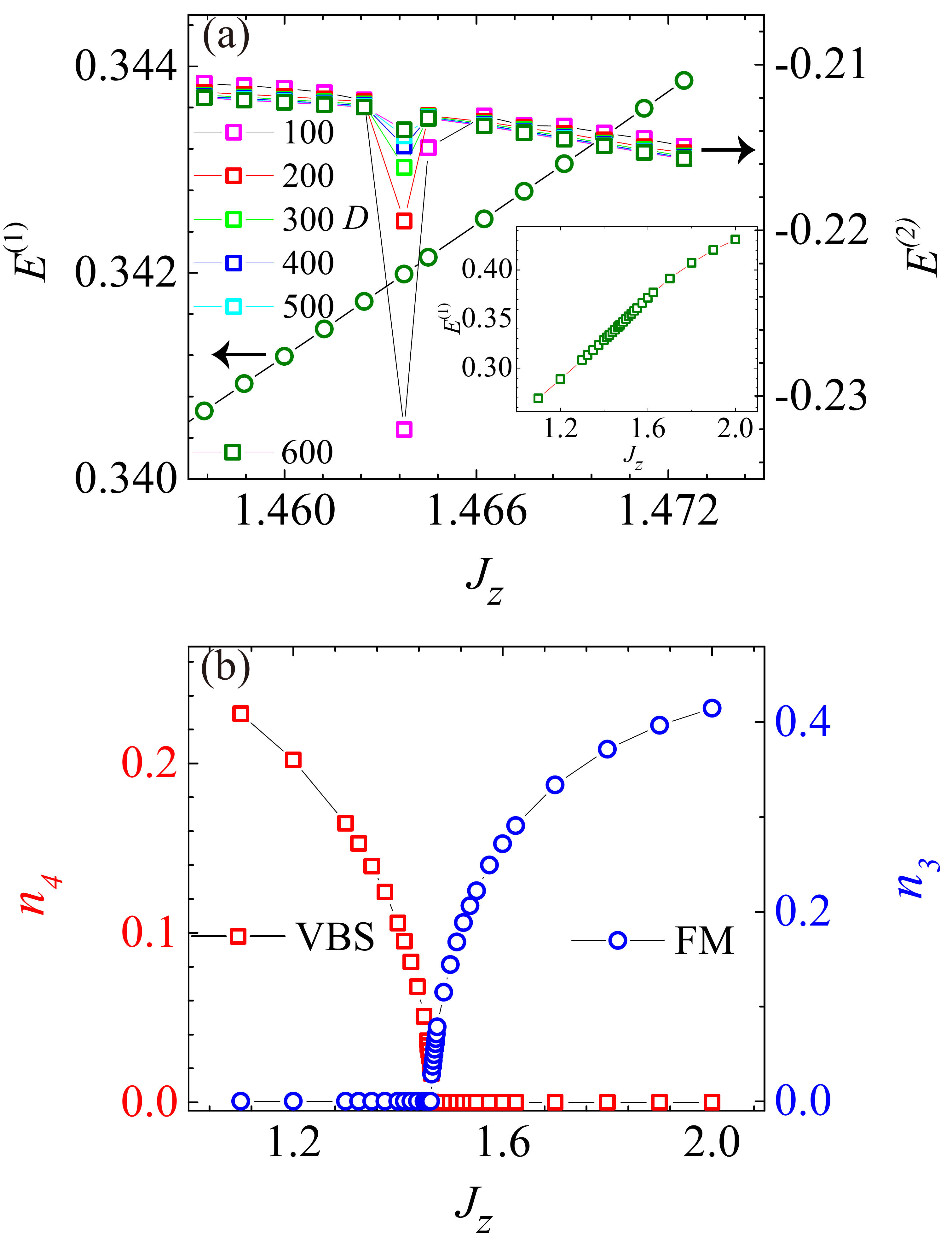}
  \caption{(a) The first and second derivatives of the ground state energy $E^{(1)}=\frac{\partial \langle H \rangle}{\partial J_z}$ and $E^{(2)}=\frac{\partial^{2}\langle H \rangle}{\partial J_z^{2}}$) with respect to $J_z$. (b) $z$-FM order parameter $n_3$ and VBS order $n_4$ as a function of $J_z$ calculated from $D=600$ MPS.}
\label{fig:orderparameter1}
\end{figure}

We present numerical results in this section.  \figref{fig:orderparameter1}(a) shows the derivatives of ground state energy with respect to the control parameter $J_z$. There is no clear discontinuity in the first derivative, while the second derivative develops a singularity at $J_c=1.4645$, the singularity decreases with increasing $D$. These results indicate that the transition is continuous with higher order.

\figref{fig:orderparameter1}(b) shows both the $z$-FM order parameter $n_3=\sum_i\langle S^{z}_{i} \rangle$ and the VBS order parameter $n_4 =\sum_i (-)^i\langle \Psi_{i} \rangle$ across the transition. Both order parameters vanish continuously at a single transition point $J_c$, demonstrating a direct and continuous DQCP. Due to the finite value of $D$ in the MPS wave function, a finite correlation length and entanglement entropy introduces a hard cut-off to the results, hence one sees a small discontinuity $\sim 0.02J_z$ to the order parameters at $J_c$. However, such discontinuity, i.e., the finite order parameter at the DQCP, is an artifact of MPS method and later we will show that such finite value of order parameters goes to zero in a power law manner with increasing $D$.

\begin{figure}[htbp]
\includegraphics[width=0.75\columnwidth]{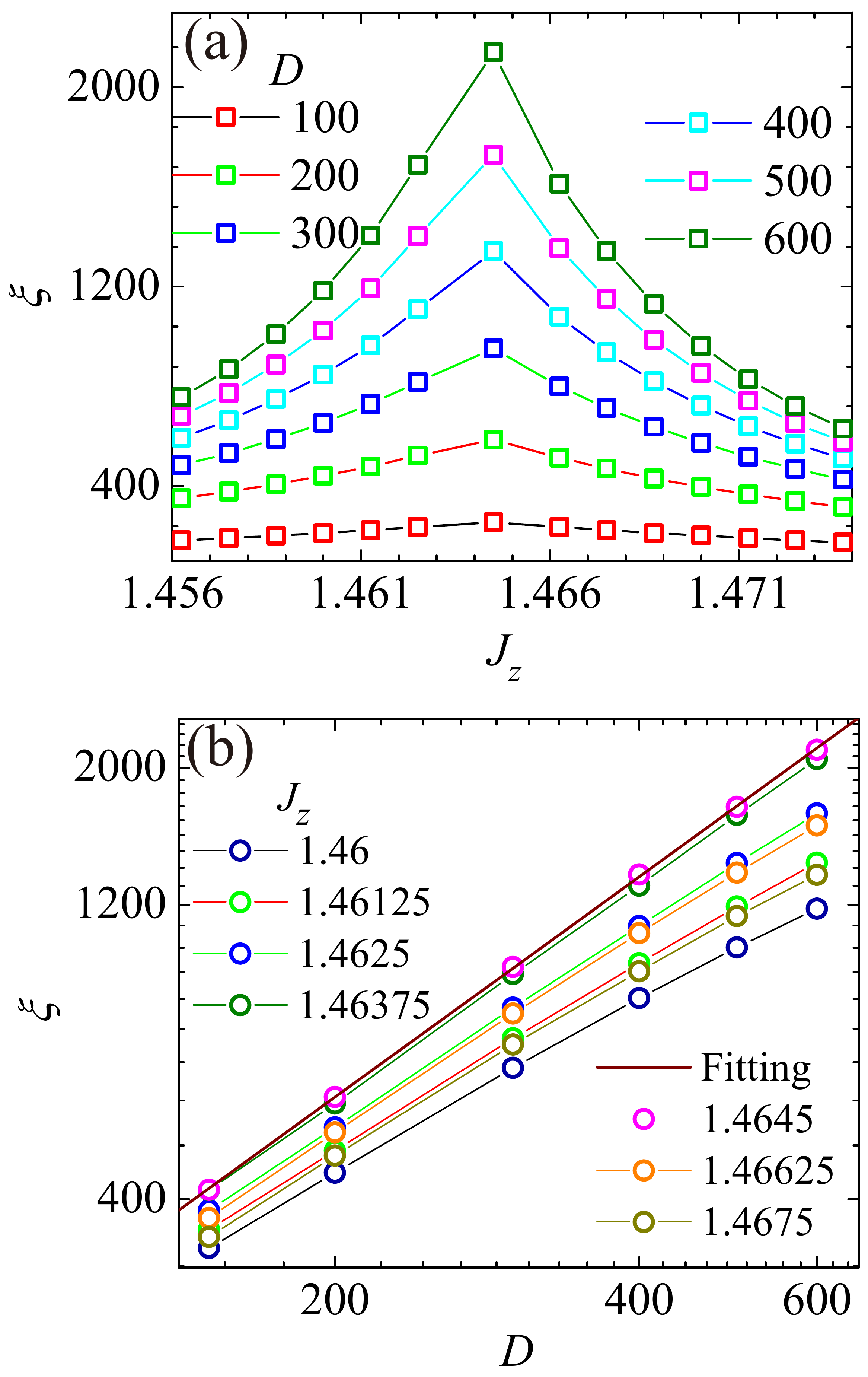}
  \caption{(a) The correlation length $\xi$ for different $J_z$ across the critical point. A divergence at $J_c=1.4645$ manifests. (b) The correlation length $\xi$ from different $D$ MPS wave function in log-log plot. Only at $J_c$, the $\xi(D)$ is a power-law function.}
\label{fig:correlation_length}
\end{figure}

To further demonstrate the criticality under different the bond dimensions $D$, we study the correlation length $\xi$, obtained from the MPS following \eqnref{eq:correlationlength}. The results are shown in \figref{fig:correlation_length}(a). It is clear that $\xi$ develops a singular behavior across $J_c$ and diverges with increasing $D$.

\begin{figure}[tbp]
\includegraphics[width=0.7\columnwidth]{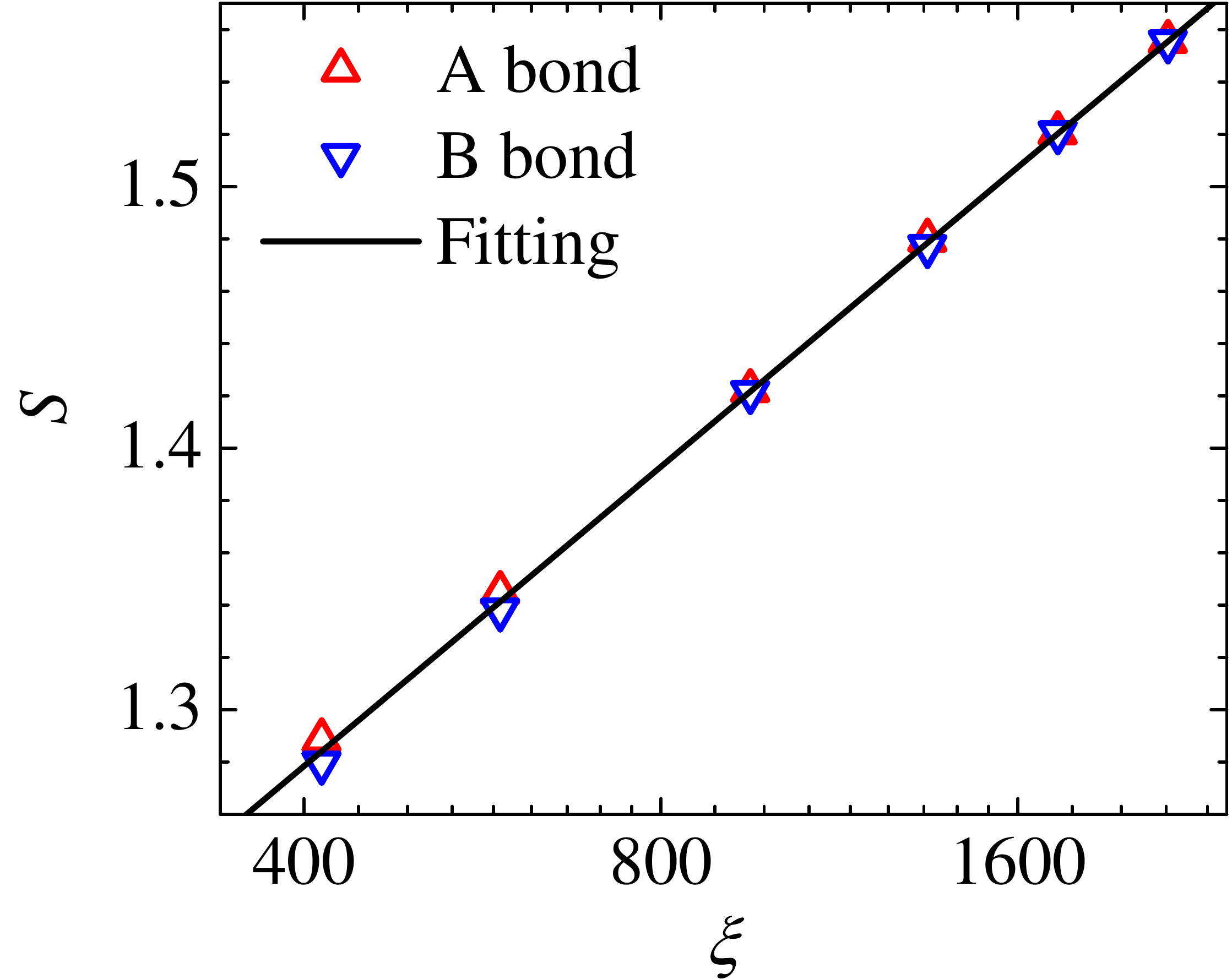}
\caption{The entanglement entropy $S$ as a function of correlation length $\xi$ at the DQPC $J_z=J_c$. The entanglement entropy at odd bond(A bond) and even bond(B bond) in the MPS are both used for the fitting, with the fitting form $S = \frac{c}{6} \log(\xi)$ and $c=0.99$ is obtained.}
\label{fig:Fig1b}
\end{figure}

Right at $J_c$, the entanglement entropy shows a perfect logarithmic function of the correlation length $\xi$ as shown in \figref{fig:Fig1b}, where by fitting the entanglement entropy $S$ as a function of the correlation length $\xi$ at $J_c$, $S = \frac{c}{6} \log(\xi)$, we obtain the central charge $c=0.99$ at the DQCP. This suggests the critical theory is a $c=1$ conformal field theory. Furthermore at $J_c$, $\xi$ shows a power law behavior $\xi(J_c)\sim D^{\kappa}$ as shown in \figref{fig:correlation_length} (b), which supports the theory of finite entanglement scaling~\cite{cite_FLS_MPS2}. The exponent $\kappa=1.18$ is close to $\frac{6}{c(\sqrt{12/c}+1)}=1.344$ as suggested in Ref.~\cite{cite_FLS_MPS2}.

\begin{figure}[htbp]
\includegraphics[width=0.75\columnwidth]{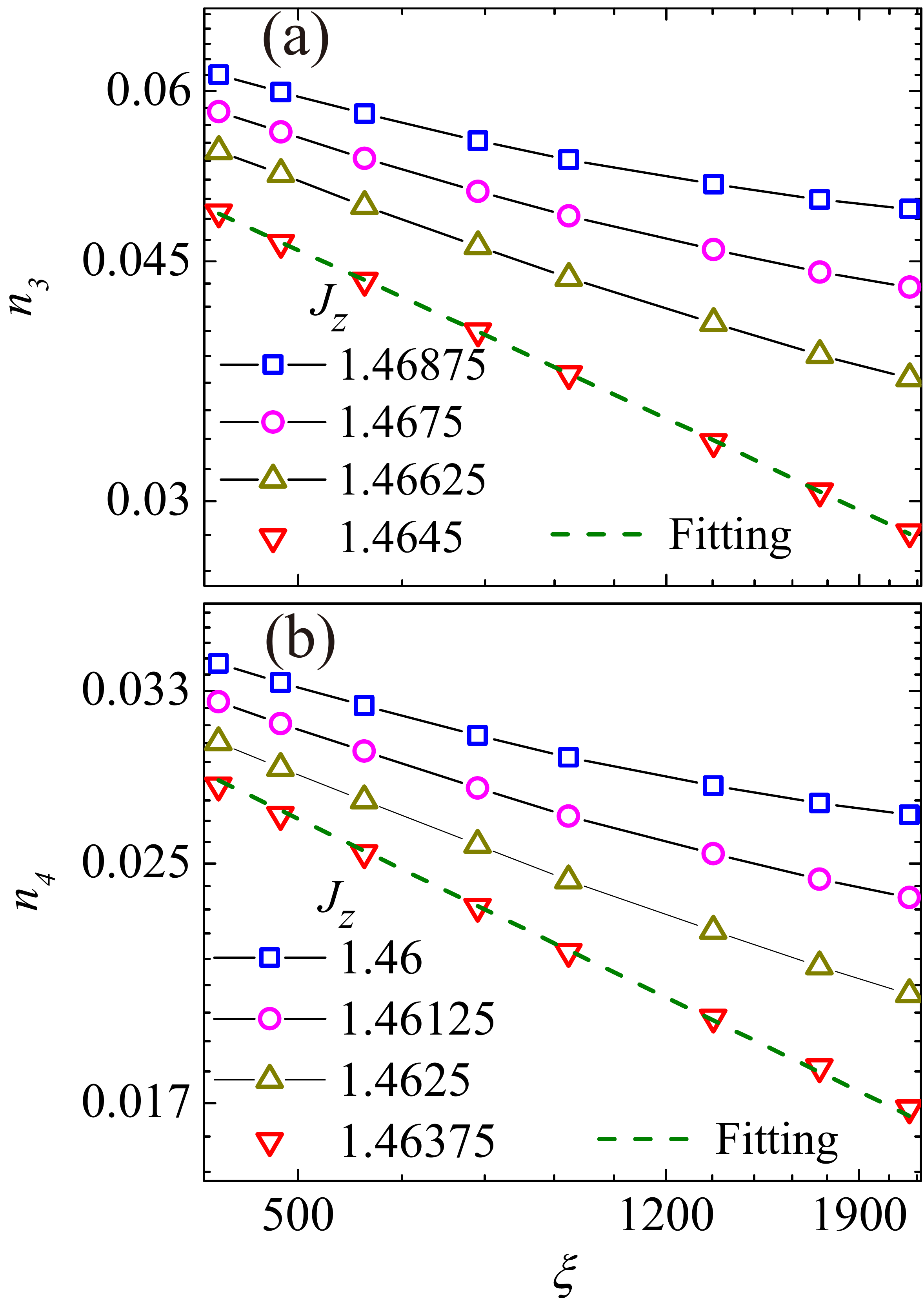}
  \caption{The log-log plot for (a) the $z$-FM order parameter $n_3$ and (b) the VBS order parameter $n_4$ as a function of correlation length $\xi$. The fittings at $J_c$ give rise to the critical exponents $\Delta_3$ and $\Delta_4$.}
\label{fig:orderparameter2}
\end{figure}

\begin{figure}[htbp]
\includegraphics[width=0.9\columnwidth]{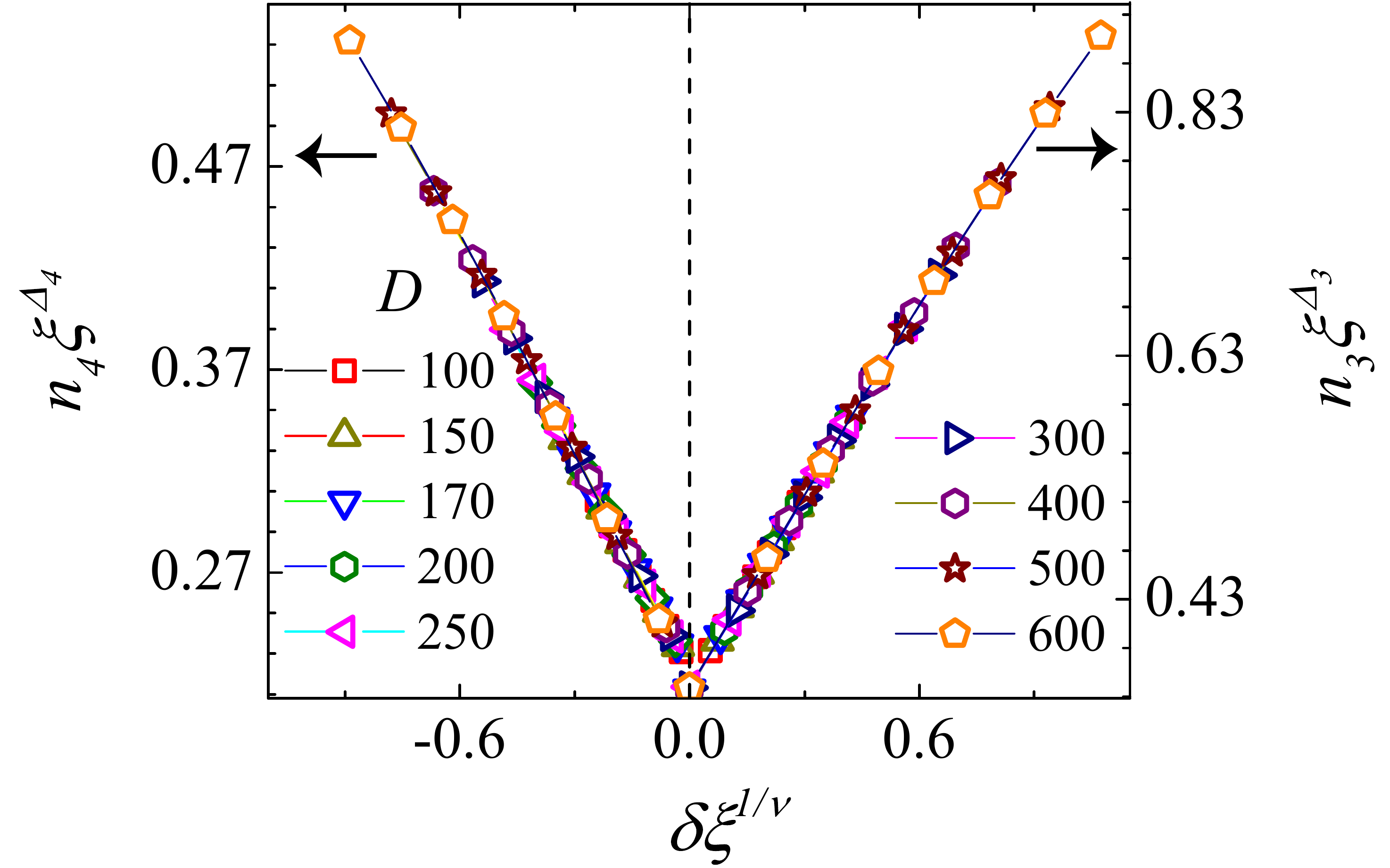}
  \caption{Data collapse of the $z$-FM order $n_3$ and the VBS order $n_4$ with correlation length $\xi$, according to Eq.\eqref{eq:Eq_scale1}. $\delta=J_z-J_c$ is the driving parameter with respect to the critical point. Critical exponents $\Delta$ and $\nu$ can be obtained from the collapse.}
\label{fig:orderparameter3}
\end{figure}

With $J_c$ determined, we can carry finite length scaling analysis to extract the critical exponents at the critical point. One can evaluate the scaling dimension $\Delta_a$ for each of the $\O(4)$ vector component $n_a$ ($a=1,2,3,4$) and the critical exponent $\nu$ independently using the scaling formula in \eqnref{eq:Eq_scale1}. As we sit right at the critical point $J_z=J_c$ ($g=0$), we have $n_a\sim \xi^{-\Delta_a}$. \figref{fig:orderparameter2} demonstrates the expected power law behavior of the leading order parameters $n_3$ ($z$-FM) and $n_4$ (VBS) at the DQCP. We found identical scaling dimensions $\Delta_3=\Delta_4=0.33$, in support of the emergent $\O(2)_\theta$ symmetry that relates $\Delta_3$ and $\Delta_4$ together. Away from the critical point, the same result can be further confirmed by collapsing the order parameters in the vicinity of DQCP using the full form of \eqnref{eq:Eq_scale1}, as shown in \figref{fig:orderparameter3} left and right panels. The data with different $D$, i.e. different $\xi$, collapse onto a single scaling function. Such data collapses can independently determine $\Delta_3=0.33(2)$ and $\Delta_4= 0.35(3)$, well consistent with those obtained in \figref{fig:orderparameter2}, and $(1/\nu)_3 = 0.62(3)$ and $(1/\nu)_4=0.61(3)$, with the subscript $3$ or $4$ labels the result obtained from collapsing the order parameter $n_3$ or $n_4$. The exponents are summarized in \tabref{tab:exponents}(a).

\begin{table}[htp]
\caption{Critical exponents measured using different methods from different channels. The Luttinger parameters $g$ calculated from all exponents consistently point to $g=1.38(1)$.}
\begin{center}
\def\arraystretch{1.4}
\setlength{\tabcolsep}{7.3pt}
\begin{tabular}{c|c|cc}
\hline\hline
method & channel & \multicolumn{2}{c}{exponents}\\
\hline
\multirow{6}{*}{\parbox{70pt}{\raggedright (a) data collapse by \eqnref{eq:Eq_scale1} around $J_c$}}
 & & $\Delta$ & $g$ \\
 \cline{2-4}
 & {$n_3$} & 0.33(2) & 1.32(8) \\
 & {$n_4$} & 0.35(3) & 1.4(1) \\
 \cline{2-4}
 &   &$1/\nu$ & $g$ \\
 \cline{2-4}
 & {$n_3$} & 0.62(3) & 1.38(3) \\
 & {$n_4$} & 0.61(3) & 1.39(3) \\
 \hline
\multirow{5}{*}{\parbox{70pt}{\raggedright (b) data collapse by \eqnref{eq:Eq_scale2} around $J_c$}}
 &  & $\eta$ & $g$ \\
 \cline{2-4}
 & $G_x^{(0)}$ & 1.45(3) & 1.38(3) \\
 & $G_y^{(\pi)}$ & 1.46(3) & 1.37(3) \\
 & $G_z^{(0)}$ & 0.68(3) & 1.36(6) \\
 & $G_\Psi^{(\pi)}$ & 0.70(2) & 1.40(4) \\
\hline
\multirow{2}{*}{\parbox{70pt}{\raggedright (c) fitting by \eqnref{eq:Eq_scale2} at $J_c$}}
 & $G_x^{(\pi)}$ & 2.10(4) & 1.46(9) \\
 & $G_y^{(0)}$ & 2.11(4) & 1.44(8) \\
 \hline
\multirow{2}{*}{\parbox{70pt}{\raggedright (d) combined  methods}}
 & $G_z^{(\pi)}$ & 2.02(6) &  \\
 & $G_\Gamma^{(\pi)}$ & 2.00(5) & \\
\hline\hline
\end{tabular}
\end{center}
\label{tab:exponents}
\end{table}


\begin{figure*}[htbp]
\includegraphics[width=\textwidth]{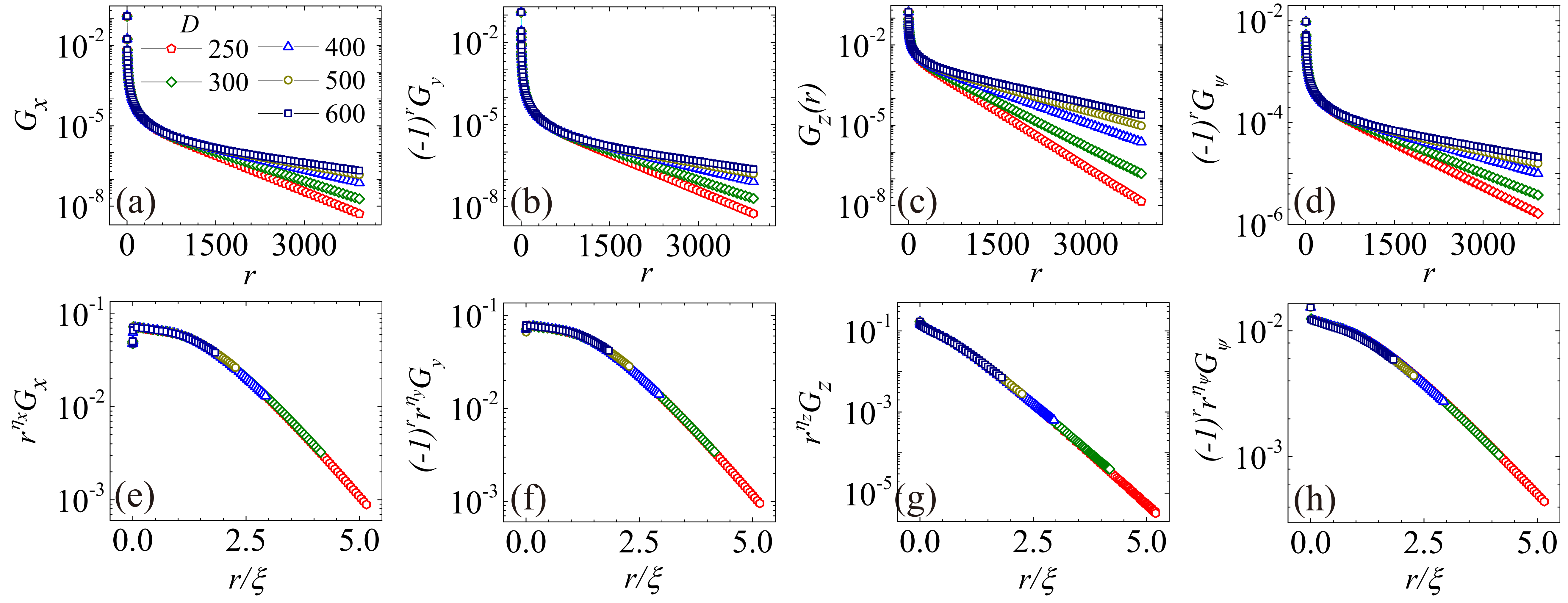}
  \caption{(a-d) The correlation functions defined in \eqnref{eq:correlations} calculated from MPSs of various $D$ at the critical point. (e-h) Data collapse of the correlations functions with different correlation lengths $\xi$. With this approach, data beyond the correlation length can also be utilized. In each channel, the exponent $\eta$ can be extracted based on \eqnref{eq:Eq_scale1}.}
\label{fig:correlation}
\end{figure*}

We then calculate the correlation functions in different channels defined in Eq.~\eqref{eq:correlations}. In general, the correlation function can be decomposed into the uniform component $G^{(0)}$ and the stagger component $G^{(\pi)}$ as ($0$ and $\pi$ label the momemtum)
\begin{equation}
G(r)=G^{(0)}(r)+(-1)^rG^{(\pi)}(r).
\end{equation}
The two components typically have different power-law exponents, so only one of them will dominate the scaling behavior at large distance. By collapsing the correlation functions $G(r)$ calculated from MPSs of different bond dimensions $D$ onto a single curve following \eqnref{eq:Eq_scale2}, as shown in \figref{fig:correlation}, the anomalous exponent $\eta$ can be extracted. Then we can determine the scaling dimension of the low-energy fluctuation that contributes to the leading component of a correlation function.
The anomalous dimension $\eta$ obtained from different correlation functions $G_x,G_y,G_z,G_\Psi$ are listed in \tabref{tab:exponents}(b), from which the Luttinger parameter $g$ can be calculated according to \eqnref{eq:luttinger}.
The Luttiger parameters evaluated from all different channels are indeed consistent with each other. Moreover, since $G_x^{(0)}$ and $G_y^{(\pi)}$ are related by the emergent $\O(2)_\phi$ symmetry, and $G_z^{(0)}$ and $G_\Psi^{(\pi)}$ are related by the emergent $\O(2)_\theta$ symmetry, the $\eta$ exponents of those symmetry related channels must be identical.  Our numerical result of the $\eta$ exponents in \tabref{tab:exponents}(b) clearly supports the emergent $\O(2)_\phi\times\O(2)_\theta$ symmetry.

\begin{figure}[htbp]
\includegraphics[width=0.7\columnwidth]{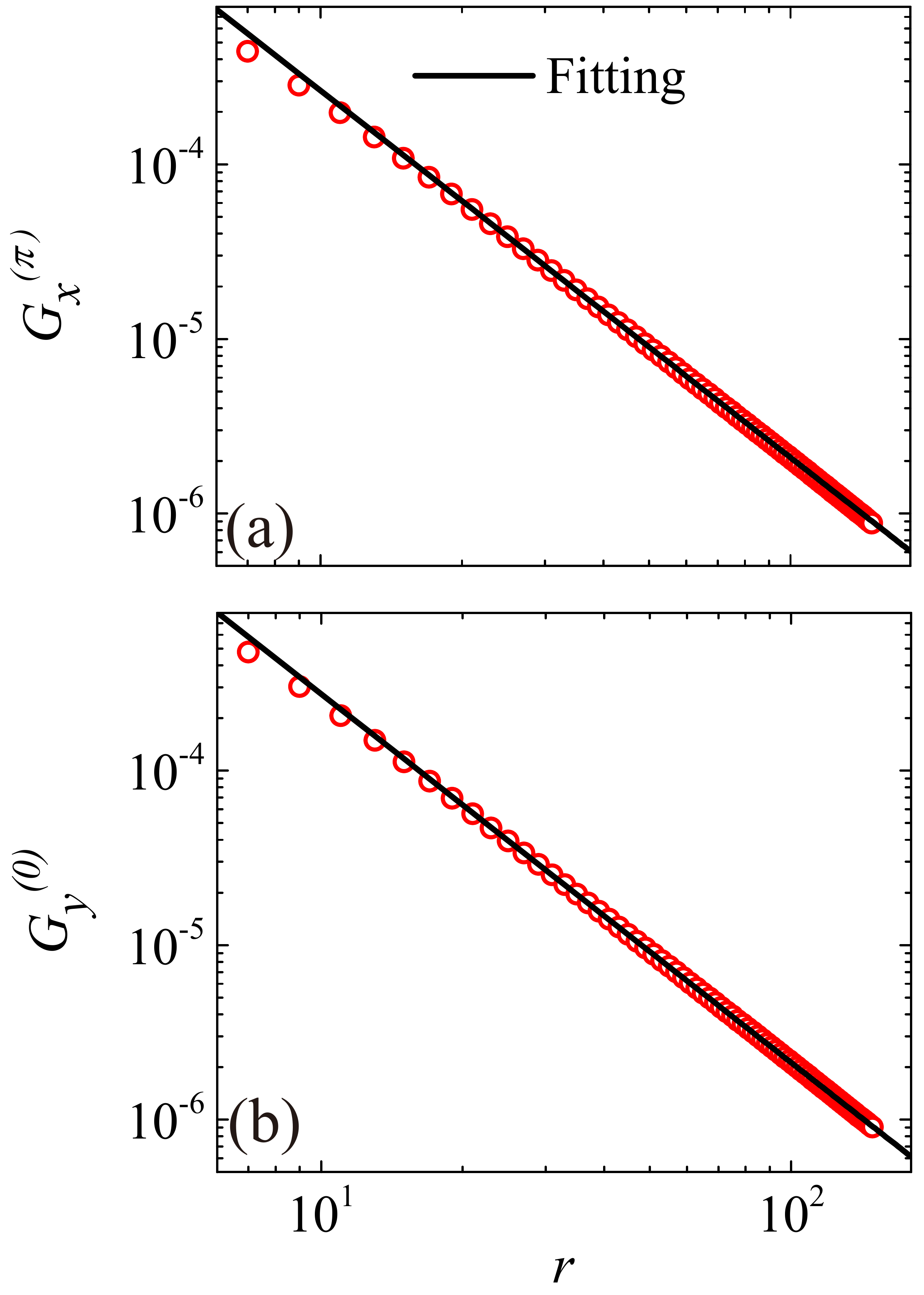}
  \caption{(a) The sub-leading component of the correlation function $G_x$ corresponds to $G_x^{(\pi)}$. (b) The sub-leading component of the correlation function $G_y$ corresponds to $G_y^{(0)}$. Their exponents $\eta$ can be extract by power-law fitting.}
\label{ref_Fig6}
\end{figure}

Having studied the leading component, we proceed with the sub-leading component of the correlation function. We need to first extract the sub-leading component from our data, as it is always overwhelmed by the leading component. We focus on the correlation function collected at the critical point, and use the fact that at short distance ($r\ll\xi$) the finite entanglement effect is very small, such that the leading component follows the power-law behavior nicely. Let us take $G_x$ for example, whose leading component is the uniform component. We first fit the correlation function $G_x(r)$ with a power-law function $C/r^\eta$ to determine the coefficient $C$, using the previously obtained exponent $\eta$. We then subtract the this leading contribution from the full correlation function as $\delta G_x(r)=G_x(r)-C/r^{\eta}$. To further remove the residue uniform component in $\delta G_x(r)$, we consider an even-odd subtraction following $G_x^{(\pi)}=\delta G_x(2r-1)-\delta G_x(2r)$ to fully expose the staggered component $G_x^{(\pi)}$. We then perform a power-law fitting of $G_x^{(\pi)}$ to determine its exponent. Similar method can be applied to $G_y$ as well, just to note that its leading component is staggered (instead of uniform), so the fitting function should be adjusted accordingly. In \figref{ref_Fig6} (a) and (b), we show that $G_x^{(\pi)}$ and $G_y^{(0)}$ are perfectly fitted by power-law functions. The obtained sub-leading exponents are listed in \tabref{tab:exponents}(c). They are almost identical due to the emergent $\O(2)_\phi$ symmetry. The Luttinger parameter determined from them are also consistent with our previous results. This completes the consistency check that different critical exponents are indeed controlled by a single Luttinger parameter as predicted in \eqnref{eq:correlations}.

\section{Emergent symmetry and conserved current}
\label{sec:Symmetry}

We have measured the sub-leading exponents of $G_x$ and $G_y$. What about the sub-leading exponent of $G_z$? It turns out that the stagger component ($z$-AFM) is sub-leading in $G_z$, which corresponds to the Noether current associated to the emergent $\O(2)\times\O(2)$ symmetry. If the continuous symmetry indeed emerges at low-energy, the conservation law will require the scaling dimension of $z$-AFM fluctuation to be pinned at $\Delta=1$ (in 1D the charge density must scale inversely with the length in order for the total charge to be conserved), such that the sub-leading exponent of $G_z$ must take $\eta=2\Delta=2$ exactly. By measuring this exponent, we can numerically determine to which degree the emergent symmetry holds.

\begin{figure}[htbp]
\includegraphics[width=\columnwidth]{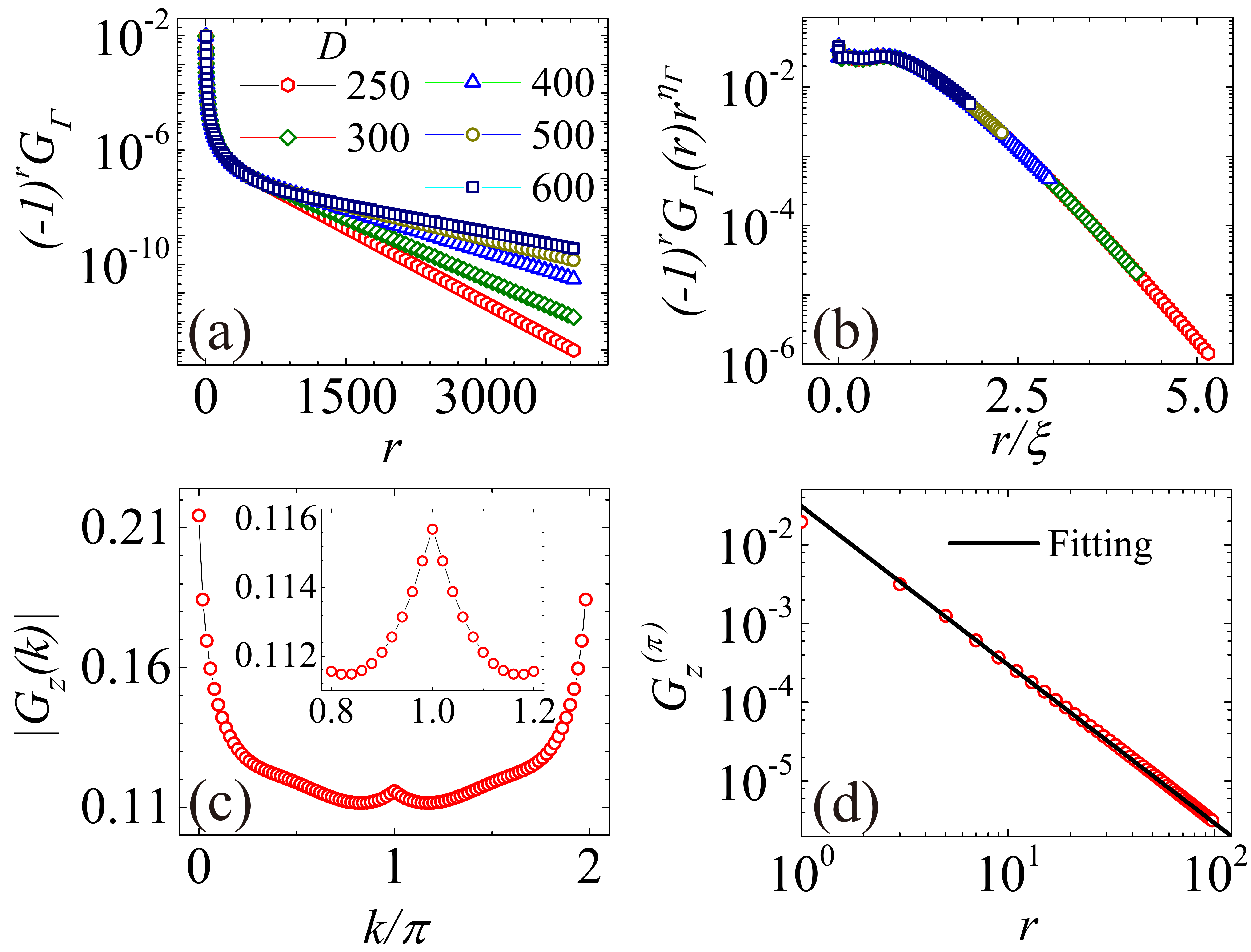}
  \caption{The correlation function of the conserved currents. (a) and (b) The correlation function of the emergent conserved current $J_x^\phi$ or $J_\tau^\theta$ before and after rescaling with correlation length $\xi$. (c) The Fourier components of the correlation function $G_z(r)$. (d) The log-log plot of the sub-leading component $G_z^{(\pi)}$.}
\label{ref_Fig5}
\end{figure}

To be more systematic, we can study all components of the conserved currents, as defined in \eqnref{eq: J=angle}, which are dictated by the emergent $\O(2)_\phi\times\O(2)_\theta$ symmetry. We first look at the current $J_x^{\phi}$ ($J_\tau^{\theta}$), which can be probed by the correlation function $G_\Gamma(r)$, according to \eqnref{eq: J=S}. The correlation function should decay in power-law with an exact exponent $2$. This is in perfectly agreement with the data presented in \figref{ref_Fig5}(a) and (b), where $G_\Gamma(r)$ determined from different $D$ MPSs precisely collapse onto each other with an exponent $2.00(5)$. Then we turn to the conserved current $J_\tau^\phi$($J_x^\phi$), which corresponds to the stagger component of $G_z$. To extract the stagger component, we first demonstrate the Fourier transform of $G_z(r)$ in \figref{ref_Fig5}(c). We observe that besides the main peak around momentum 0 (or $2\pi$), there is also a small peak representing the sub-leading component at momentum $\pi$. Using the same method of measuring the sub-leading scaling in the correlation function in Sec.~\ref{sec:FLS}, we can extract $G_z^{(\pi)}$ and perform the power-law fitting in \figref{ref_Fig5}(d). The optimal fitting parameter gives an exponent $2.02(6)$. As summarized in \tabref{tab:exponents}(d), both the exponents of $xy$-VBS and $z$-AFM are very close the exact exponent $2$, indicating the conservation of the emergent currents for both $\O(2)$ symmetries. Our results strongly support the emergent $\O(2)_\phi \times \O(2)_\theta$ symmetry predicted by the anisotropic $\O(4)$ NLSM field theory and also shows the exponents evaluated in the finite length scaling are reliable.

\section{Summary}
\label{sec:Summary}
In this work, we develop an anisotropic $\O(4)$ NLSM field theory for a 1D incarnation of deconfined quantum critical point between two $\dsZ_2$ breaking phases (which may be called an Ising-DQCP). The theory provides the scaling laws of various correlation functions at the Ising-DQCP and they are all governed by a single Luttinger parameter. Moreover, we confirm the emergent continuous symmetries O(2)$_\phi$ $\times$ O(2)$_{\theta}$ by measuring the associated conserved currents. The emergent conserved current fluctuation can be considered as a hallmark of the DQCP, which is not expected for a conventional Ising transition.

We systematically employed variational MPS method to study the proposed properties of this Ising-DQCP. The phase transition and the critical point is determined from the singular behavior of the ground state energy, order parameter and correlation length. By means of finite length scaling we obtain the critical exponents independently. The exponents such obtained are consistent with each other and the Luttinger liquid theory. Furthermore we confirm the emergent symmetry using the conserved current and find the scaling dimension for the conserved current is consistent with the NLSM field theory. At the numerical level, this work can be viewed as the first test of the finite length scaling in the MPS representation upon a non-trivial DQCP with unknown exponents, in this regard, our systematic analysis sets a protocol for future research in more exotic systems with MPS and tensor network methods.

Our work provides a solid example of low-dimensional analog of DQCP where unbiased numerical results are in perfect agreement with the controlled field theory predictions, and has extended the realm of the DQCP as well as its discovery with advanced numerical methodology one step further. It is worth mentioning that \eqnref{eq:model} is not the only example of 1D analog of DQCP. Already in quantum spin chain systems~\cite{AffleckHaldane1987,Patil2018,Mudry2019} emergent continuous symmetry has been addressed analytical and numerically. And 1D extended Hubbard model~\cite{Sengupta2002,Sandvik2004} is studied with quantum Monte Carlo and a direct continuous transition between a charge density wave (CDW) phase and a bond order wave (BOW) phase is found. The CDW and BOW phases respectively break the bond-centered and site-centered reflection symmetries. The lattice symmetries together with the charge $\U(1)$ symmetry also lead to a generalized LSM constraint\cite{Cho2017} that ensures the transition to be critical. So the CDW-BOW transition also qualifies as a 1D analog of DQCP. At the time of Refs.~\cite{Sengupta2002,Sandvik2004}, it was not realized that there exists an emergent $\O(2)$ symmetry at the critical point, rotating the CDW and BOW order parameters. The emergent conserved current corresponds to the charge density fluctuation. This implies that there are many low-dimensional systems acquiring phase transitions beyond LGW paradigms and our study here could motivate further research of the emergent conserved current in such systems.

{\it Note added:} We would like to draw the reader's attention to a related parallel work by Brenden Roberts, Shenghan Jiang and Olexei Motrunich to appear in the same arXiv posting.

\section*{Acknowledgments}
R.Z.H. thanks Shuai Yin for suggesting this interesting question and stimulating discussions about the scaling analysis, Hai-Jun Liao for useful discussion about the numerical calculation of the ground state wave function. Y.Z.Y. thanks the insightful discussion with Chao-Ming Jian, Meng Cheng, Zhen Bi and Ashvin Vishwanath on generalized LSM theorems. We also thank Bruce Normand for helpful discussion and Anders W. Sandvik for introducing several previous literatures in which 1D analog of DQCP is already presented. Z.Y.M. thanks Snir Gazit and the Racah Institute of Physics in the Hebrew University of Jerusalem for providing encouraging working atmosphere, in which this work is finalized. We acknowledge the support from the Strategic Priority Research Program of Chinese Academy of Sciences (Grant XDB28000000), the National Science Foundation of China (Grant NSFC-11674278 and 11574359) and the National Key R\&D Program (Grant-2016YFA0300502).

\bibliographystyle{apsrev4-1}
\bibliography{references}

\appendix
\onecolumngrid
\section{Fermionic Spinon Theory}\label{sec: fermionic}

\subsection{Fractionalization Scheme}\label{sec:fractionalization}

We propose a fermionic spinon theory for the $\O(4)$ non-linear $\sigma$-model (NLSM) that describes the $z$-FM--VBS transition. The fermionic spinon theory allows us to make direct connections between the microscopic lattice spin operators and the field theory operators, which enable us to identify the spin operator representation of the emergent conserved currents at the critical point.

The fermionic spinon theory starts with introducing the fermionic spinon $f_i=(f_{i\uparrow},f_{i\downarrow})^\intercal$ on each site $i$, where $f_{i\uparrow}$ and $f_{i\downarrow}$ are complex fermion operators. The spin operator on each site is fractionalized as follow,
\eq{\label{eq:Si}S_i^x=(-)^i f_i^\dagger\sigma^x f_i,\quad S_i^y= f_i^\dagger\sigma^y f_i,\quad S_i^z=(-)^i f_i^\dagger\sigma^z f_i.}
The stagger factor $(-)^i$ makes the spins alternate between the sublattices. They are assigned in such way that all the relevant spin orders $x$-FM, $y$-AFM and $z$-FM are unified as a single $\O(3)$ vector, which is the AFM order in terms of the fermionic spinons. Let us place the fermionic spinons in the Dirac band structure described by the following mean-field Hamiltonian,
\eq{\label{eq:HMF}H_0=-\frac{1}{2}\sum_{i}\ii f_i^\dagger f_{i+1}+\text{h.c.}.}
Note that this fermion mode \eqnref{eq:HMF} can \emph{not} be derived from the original lattice spin model by plugging in \eqnref{eq:Si}. \eqnref{eq:HMF} should rather be considered as a parton construction for the $\O(4)$ NLSM, which only describes the vicinity of the critical point. We can further establish the following operator correspondence,
\eq{\label{eq:DiGammai}\Psi_i=\vect{S}_i\cdot \vect{S}_{i+1}\sim -(\ii f_i^\dagger f_{i+1}+\text{h.c.})/2,\quad \Gamma_i=S_i^x S_{i+1}^y+S_i^y S_{i+1}^x\sim (f_i^\dagger\sigma^z f_{i+1}+\text{h.c.})/2,}
where $\Psi_i$ is the diagonal dimmer operator and $\Gamma_i$ is the off-diagonal dimmer operator. One may wonder that the dimmer operators should contain four fermions according to \eqnref{eq:Si}, but the fact is that under RG flow, the four-fermion operators will generate the above two-fermion operators, which consitute the most relevant component of the dimmer operators. The identification will be justified by the symmetry analysis later. The physical picture is that the dimmer is a bond ordering that modulates the bond energy. In the spinon Hamiltonian \eqnref{eq:HMF}, the only way to modulate the bond energy is to change the hopping strength, so the hopping term should be identified as the dimmer operator. The difference between the diagonal and off-diagonal dimmer is that the former one($\Psi_i$) does not break the spin reflections $g_x$ and $g_z$ and is therefore the ordinary hopping, while the later ($\Gamma_i$) also is odd under $g_x$ and even under $g_z$, like $S^x S^y\sim S^z$ and should be represented as $\sigma^z$ spin-dependent hopping.

We take a two-site unit cell and label the sublattices by $A$ and $B$ respectively. The Hamiltonian can be diagonalized in momentum space, introducing the basis in momentum space,
\eq{f_{k}=\smat{f_{kA}\\f_{kB}}\otimes\smat{\uparrow\\\downarrow}=(f_{kA\uparrow},f_{kA\downarrow},f_{kB\uparrow},f_{kB\downarrow})^\intercal,}
and rewrite the mean-field Hamiltonian \eqnref{eq:HMF} in the momentum space as,
\eq{H_0=\sum_{k}f_{k}^\dagger (\sin k)\sigma^{10} f_k,}
where $\sigma^{ab}=\sigma^a\otimes\sigma^b$ denotes the tensor product of Pauli matrices. The first Pauli matrix acts on the sublattice space and the second Pauli matrix acts on the spin space. As one unit-cell contains two sites, the momentum takes values in the Brillouin zone $[-\pi/2,\pi/2)$. The low-energy fermions are around the $k=0$ point. Expanding the Hamiltonian around the momentum $k=0$ point, and we introduce the fermionic spinon field $f\sim f_{k\to 0}$. The mean field Hamiltonian in the long wave length limit can be written as,
\eq{\label{eq:H0}H_0=\int\dd x f^\dagger (-\ii\partial_x \sigma^{10})f,}
where $f=(f_{A\uparrow},f_{A\downarrow},f_{B\uparrow},f_{B\downarrow})^\intercal$ is a four-component complex fermion field describing the low-energy spinon. The real space operators can be transformed to the momentum space as,
\eq{\vect{S}_Q=\sum_i \vect{S}_i e^{-\ii Q i},\quad \Psi_Q=\sum_i \Psi_i e^{-\ii Q i},\quad \Gamma_Q=\sum_i \Gamma_i e^{-\ii Q i}.}
Their representations in terms of the fermionic spinon field are concluded in \tabref{tab:ops}.

\begin{table}[htp]
\caption{Spin and dimmer operators in terms of fermionic spinon field bilinears}
\begin{center}
\begin{tabular}{ccc}
& $Q=0$ & $Q=\pi$\\
\hline
$S^x_Q$ & $f^\dagger \sigma^{31}f$ & $f^\dagger \sigma^{01}f$\\
$S^y_Q$ & $f^\dagger \sigma^{02}f$ & $f^\dagger \sigma^{32}f$\\
$S^z_Q$ & $f^\dagger \sigma^{33}f$ & $f^\dagger \sigma^{03}f$\\
$\Psi_Q$ & $f^\dagger \sigma^{10}f$ & $f^\dagger \sigma^{20}f$\\
$\Gamma_Q$ & $f^\dagger \sigma^{23}f$ & $-f^\dagger \sigma^{13}f$
\end{tabular}
\end{center}
\label{tab:ops}
\end{table}

By definition, the orders $x$-FM ($S_0^x$), $y$-AFM ($S_\pi^y$), $z$-FM ($S_0^z$) and VBS ($\Psi_\pi$) forms an $\O(4)$ vector. So the  $\O(4)$ vector $\vect{n}=(n_1,n_2,n_3,n_4)$ couples to the fermionic spinon by,
\eq{n_1 S_0^x+n_2 S_\pi^y+n_3 S_0^z+n_4 \Psi_\pi=f^\dagger(n_1\sigma^{31}+n_2 \sigma^{32}+n_3 \sigma^{33}+n_4 \sigma^{20})f.}
The relation between the spinon mean-field theory and the $\O(4)$ NLSM is now clarified by the fermion $\sigma$-model (FSM), which extends the spinon field Hamiltonian in \eqnref{eq:H0} to include the interaction with the $\O(4)$ vector field
\eq{\label{eq:FSM}H_\text{FSM}=\int \dd x f^\dagger(-\ii\partial_x \sigma^{10}+n_1\sigma^{31}+n_2 \sigma^{32}+n_3 \sigma^{33}+n_4 \sigma^{20})f.}
If we integrate out the fermions in \eqnref{eq:FSM}, we arrived at the effective theory for the $\O(4)$ vector field $\vect{n}$ described by the following action,
\eq{\label{eq:NLSM0}S_\text{NLSM}=\int\dd^2x \frac{1}{2\kappa}(\partial_\mu\vect{n})^2
+\frac{\ii k}{2\pi^2} \epsilon^{abcd}n_a\partial_\tau n_b\partial_x n_c\partial_u n_d,}
which is the non-linear $\sigma$ model (NLSM) with WZW term at level $k=1$ and $\kappa$ is a non-universal constant describing the stiffness of the $\O(4)$ vector field generated by the dynamics of the fermionic spinon. The Dirac fermion in \eqnref{eq:FSM} provides a physical mechanism to generate the WZW term in \eqnref{eq:NLSM0}.

However, the FSM in \eqnref{eq:FSM} has a larger Hilbert space than the NLSM in \eqnref{eq:NLSM0}. For example, the spectrum of FSM contains fermionic excitations that are not present in the spectrum of NLSM, so the two theories are not equivalent. In fact, at low energy, the FSM flows to a $\U(2)_1\simeq\O(4)_1$ CFT, while the NLSM flows to a $\SU(2)_1$ CFT. However, the two theories are related to each other. In each chiral sector, we have the following CFT decomposition:
\eq{\U(2)_1\simeq\O(4)_1\simeq\SU(2)_1 \oplus \SU(2)_1,}
meaning that the $\U(2)_1$ or the $\O(4)_1$ CFT, which describes 2 copies of free complex fermions or 4 copies of free Majorana fermions, can be decomposed into the direct sum of two interacting $\SU(2)_1$ CFTs. The two $\SU(2)$ groups can be separately interpreted as the spin $\SU(2)$ and the gauge $\SU(2)$. The gauge $\SU(2)$ structure arises from fractionalizing the spin operator into fermionic spinons, as the following gauge transformation on the spinon field does not have any physical consequence ($H_0$ and all the spin/dimer operators remain invariant),
\eq{\mat{f_{i\uparrow}\\f_{i\downarrow}^{\dagger}}\to U_i\mat{f_{i\uparrow}\\f_{i\downarrow}^{\dagger}},}
for any $\SU(2)$ matrix $U_i$ on every site $i$. In the spinon language, the gauge $\SU(2)$ generator on site $i$ is given by
\eq{\vect{q}_i=(\Re f_i^\intercal \ii\sigma^{y} f_i,\Im f_i^\intercal \ii\sigma^{y} f_i,f_i^\dagger f_i-1),}
where $\Re$ and $\Im$ act on an generic operator $\scO$ as $\Re \scO=(\scO+\scO^\dagger)/2$ and $\Im \scO=(\scO-\scO^\dagger)/(2\ii)$. In terms of the spinon field at low-energy, one can write down the gauge charge $\vect{J}_0$ and current $\vect{J}_1$,
\eq{\vect{J}_0=(\Re f^\intercal \ii\sigma^{02}f,\Im f^\intercal \ii\sigma^{02}f,f^\dagger \sigma^{00}f),\quad \vect{J}_1=(\Re f^\intercal \ii\sigma^{12}f,\Im f^\intercal \ii\sigma^{12}f,f^\dagger \sigma^{10}f).}
To restrict the FSM to the NLSM, we must impose the gauge constraint $\vect{q}_i=0$ on every lattice site $i$ to project the FSM Hilbert space to the gauge neutral sector. At low-energy, the gauge constraint can be effectively implemented by applying a large energy penalty $U\gg 1$ to suppress the $\SU(2)$ gauge charge and current fluctuations,
\eq{H_\text{int}=\frac{U}{4}\int\dd x (\vect{J}_0\cdot \vect{J}_0+\vect{J}_1\cdot\vect{J}_1)=U\int\dd x\vect{J}_L\cdot \vect{J}_R,}
where $\vect{J}_L=(\vect{J}_0-\vect{J}_1)/2$ and $\vect{J}_R=(\vect{J}_0+\vect{J}_1)/2$ are introduced to denote the left- and right-moving gauge $\SU(2)$ currents in the CFT. $H_\text{int}$ is a back-scattering term of the gauge currents, which gaps out the gauge $\SU(2)_1$ CFT below the energy scale of $U$, leaving the spin $\SU(2)_1$ CFT at low-energy. The mechanism will be analyzed in detail later. Our proposal is that the following interacting fermion spinon field theory provides a dual description of the $\O(4)$ NLSM with $k=1$ WZW term,
\eq{\label{eq:Hf}H_f=H_0+H_\text{int}=\int\dd x f^\dagger(-\ii\partial_x\sigma^{10})f+U\vect{J}_L\cdot\vect{J}_R.}
The fermionic description will enable us to derive the emergent conserved current and to develop a more tractable bosonization theory.

\subsection{Symmetry Analysis}\label{sec:symmetry}

The fermionic spinon model facilitates us to match the symmetry between operators in microscopic model and field theory. Here the symmetries in consideration include the lattice translation symmetry $T_x$, the $\dsZ_2^x\times\dsZ_2^z$ spin flip symmetries $g_x$ and $g_z$, the time-reversal symmetry $\scT$ and the site-centered spacial-reflection symmetry $\scP$. These symmetries are defined by their action on the lattice spin operator $\vect{S}_i$ as summarized in \tabref{tab:symm}(a), with two additional rules that the time-reversal also change the sign of imaginary unit $\scT:\ii\to-\ii$ (denoted by the complex conjugation operator $\scK$) and the spacial-reflection also flips the direction of space $\scP:x\to-x$. In the following, we will establish the rest part of \tabref{tab:symm} for the symmetry transformations of different fields and operators step by step.

\begin{table}[htp]
\caption{Symmetry transformations of fields and operators.}
\begin{center}
\begin{tabular}{cc|ccccc}
& field & $T_x$ & $g_x$ & $g_z$ & $\scT$ & $\scP$\\
\cline{2-7}
\multirow{3}{*}{(a)}
& $S_i^x$ & $S_{i+1}^x$ & $S_i^x$ & $-S_i^x$ & $-S_i^x$ & $S_{-i}^x$\\
& $S_i^y$ & $S_{i+1}^y$ & $-S_i^y$ & $-S_i^y$ & $-S_i^y$ & $S_{-i}^y$\\
& $S_i^z$ & $S_{i+1}^z$ & $-S_i^z$ & $S_i^z$ & $-S_i^z$ & $S_{-i}^z$\\
\cline{2-7}
\multirow{2}{*}{(b)}
& $f_i$ & $\sigma^{y}f_{i+1}$ & $\sigma^{x}f_i$ & $\sigma^{z}f_i$ & $\scK (-)^i f_i^\dagger$ & $(-)^i f_{-i}$\\
& $f$ & $\sigma^{12}f$ & $\sigma^{01}f$ & $\sigma^{03}f$ & $\scK\sigma^{30}f^\dagger$ & $\sigma^{30}f$\\
\cline{2-7}
\multirow{4}{*}{(c)}
& $n_1$ & $n_1$ & $n_1$ & $-n_1$ & $-n_1$ & $n_1$\\
& $n_2$ & $-n_2$ & $-n_2$ & $-n_2$ & $-n_2$ & $n_2$\\
& $n_3$ & $n_3$ & $-n_3$ & $n_3$ & $-n_3$ & $n_3$\\
& $n_4$ & $-n_4$ & $n_4$ & $n_4$ & $n_4$ & $-n_4$\\
\cline{2-7}
\multirow{2}{*}{(d)}
& $\phi$ & $-\phi$ & $-\phi$ & $\phi+\pi$ & $\phi+\pi$ & $\phi$\\
& $\theta$ & $-\theta$ & $-\theta+\pi$ & $\theta$ & $-\theta+\pi$ & $-\theta$\\
\cline{2-7}
\multirow{4}{*}{(e)}
& $J_\tau^\phi$ & $-J_\tau^\phi$ & $-J_\tau^\phi$ & $J_\tau^\phi$ & $-J_\tau^\phi$ & $J_\tau^\phi$\\
& $J_x^\phi$ & $-J_x^\phi$ & $-J_x^\phi$ & $J_x^\phi$ & $J_x^\phi$ & $-J_x^\phi$\\
& $J_\tau^\theta$ & $-J_\tau^\theta$ & $-J_\tau^\theta$ & $J_\tau^\theta$ & $J_\tau^\theta$ & $-J_\tau^\theta$\\
& $J_x^\theta$ & $-J_x^\theta$ & $-J_x^\theta$ & $J_x^\theta$ & $-J_x^\theta$ & $J_x^\theta$\\
\cline{2-7}
\multirow{2}{*}{(f)}
& $S_{\pi}^z$ & $-S_{\pi}^z$ & $-S_{\pi}^z$ & $S_{\pi}^z$ & $-S_{\pi}^z$ & $S_{\pi}^z$\\
& $\Gamma_{\pi}$ & $-\Gamma_{\pi}$ & $-\Gamma_{\pi}$ & $\Gamma_{\pi}$ & $\Gamma_{\pi}$ & $-\Gamma_{\pi}$ \\
\cline{2-7}
\multirow{2}{*}{(g)}
& $\psi_L$ & $-\sigma^y\psi_L$ & $\sigma^x\psi_L$ & $\sigma^z\psi_L$ & $\scK\psi_R^\dagger$ & $\psi_R$\\
& $\psi_R$ & $\sigma^y\psi_R$ & $\sigma^x\psi_R$ & $\sigma^z\psi_R$ & $\scK\psi_L^\dagger$ & $\psi_L$\\
\end{tabular}
\end{center}
\label{tab:symm}
\end{table}%

According to the fractionalization scheme in \eqnref{eq:Si}, one can infer the symmetry transformations of the spinons, as enumerated in \tabref{tab:symm}(b). Take the translation symmetry $T_x:\vect{S}_i\to\vect{S}_{i+1}$ for example. Plugging in \eqnref{eq:Si}, the transformation rule requires the following to hold,
\eq{T_x: f_i^\dagger\sigma^x f_i\to -f_{i+1}^\dagger\sigma^x f_{i+1}, f_i^\dagger\sigma^y f_i\to f_{i+1}^\dagger\sigma^y f_{i+1}, f_i^\dagger\sigma^z f_i\to -f_{i+1}^\dagger\sigma^z f_{i+1}, }
where the $\sigma^x$ and $\sigma^z$ terms acquire the $(-)$ signs due to the stagger factor $(-)^i$ in the fractionalization scheme. To produce these signs, we must perform $\sigma^y$ rotation to the spinon as we translate it, thus $T_x:f_i\to \sigma^y f_{i+1}$. As we switch to the momentum space, the sublattice freedom is absorbed to the spinon field $f$. The translation exchanges the sublattice, which is implemented by $\sigma^1$ in the sublattice space, so $T_x:f\to\sigma^{12}f$ for the spinon field. Following the similar approach, the transformation of the spinon field under the other symmetries can be verified. It worth mention that under $\scT$ and $\scP$, the spinon transforms with the additional stagger sign $(-)^i$. This is actually a gauge choice to ensure that the mean-field Hamiltonian $H_0$ remains invariant. Because the spinon is not a gauge neutral object, symmetries are allow to fractionalize on the spinon. The non-trivial symmetry fractionalization patterns are as follows,
\eq{T_xg_x=-g_xT_x, T_xg_z=-g_zT_x, T_x\scT=-\scT T_x, T_x\scP=-\scP T_x, g_x g_z=-g_z g_x.}
Applying the symmetry transformation of the spinon field $f$, we can further determine that of the $\O(4)$ vector field $\vect{n}$, by requiring $H_\text{FSM}$ to be invariant. The results are summarized in \tabref{tab:symm}(c). We gave two examples to demonstrate the calculation method:
\eqs{n_4\sim f^\dagger \sigma^{20}f &\overset{T_x}{\to}f^\dagger\sigma^{12}\sigma^{20}\sigma^{12}f =-f^\dagger \sigma^{20} f\sim -n_4,\\
n_2\sim f^\dagger \sigma^{32}f &\overset{\scT}{\to}f\sigma^{30}\scK\sigma^{32}\scK\sigma^{30}f^\dagger =-f^\dagger(\sigma^{30}(\sigma^{32})^*\sigma^{30})^\intercal f=-f^\dagger \sigma^{32} f\sim -n_2.}

Now we turn to the continuous symmetry in the model. The FSM \eqnref{eq:FSM} and the NLSM \eqnref{eq:NLSM0} both preserves the $\O(4)$ symmetry that rotates the $\O(4)$ vector $\vect{n}$. But they are not the final theory that describe the $z$-FM--VBS transition. To describe the transition, we need to add anisotropy term $\mu (n_1^2+n_2^2-n_3^2-n_4^2)$ to the theory, which breaks the $\O(4)$ group to its $\O(2)_\phi\times\O(2)_\theta$ subgroup. The meaning of the $\O(2)_\phi\times\O(2)_\theta$ symmetry is clear if we define
\eq{\label{eq:n+in}e^{\ii\phi}\sim n_1+\ii n_2, e^{\ii\theta}\sim n_3+\ii n_4.}
Then $\O(2)_\phi$ is the orthogonal transformation of $(n_1,n_2)$ that contains a $\U(1)_\phi$ subgroup of the $\phi$-rotation. Similarly $\O(2)_\theta$ is the orthogonal transformation of $(n_3,n_4)$ that contains a $\U(1)_\theta$ subgroup of the $\theta$-rotation. In terms of the infinitesimal transformation, we have
\eqs{\label{eq:U(1)transform}&\U(1)_\phi:\phi\to\phi+\dd\phi,\mat{n_1\\n_2}\to\mat{1&-\dd\phi\\ \dd\phi & 1}\mat{n_1\\n_2},f\to(1-\tfrac{\ii}{2}\sigma^{03}\dd\phi)f;\\
&\U(1)_\theta:\theta\to\theta+\dd\theta,\mat{n_3\\n_4}\to\mat{1&-\dd\theta\\ \dd\theta & 1}\mat{n_3\\n_4},f\to(1+\tfrac{\ii}{2}\sigma^{13}\dd\theta)f.}
The transformation for the spinon field $f$ can be verified as following (take $\U(1)_\theta$ transformfor instance)
\eqs{e^{\ii\theta}=n_3+\ii n_4\sim f^\dagger(\sigma^{33}+\ii\sigma^{20})f
\overset{\U(1)_\theta}{\to}&
f^\dagger(1-\tfrac{\ii}{2}\sigma^{13}\dd\theta)(\sigma^{33}+\ii\sigma^{20})(1+\tfrac{\ii}{2}\sigma^{13}\dd\theta)f\\
&=f^\dagger((\sigma^{33}+\ii\sigma^{20})-(\sigma^{20}-\ii\sigma^{33})\dd\theta+\cdots)f\\
&\sim (n_3+\ii n_4)-(n_4-\ii n_3)\dd\theta=e^{\ii(\theta+\dd\theta)}.}

As argued in the main text, the emergent symmetry at the critical point is $\U(1)_\phi\times\U(1)_\theta$. We will provide more detailed evidence for the emergent symmetry, but let us accept this fact for now and apply the Noether theorem to calculate the corresponding emergent conserved current. For the purpose of identifying the conserved current operator, we can ignore the interaction of the spinon for a moment, because the conserved current operator does not receive renormalization from interaction as long as symmetry is preserved. Thus we start with the spinon field Hamiltonian in \eqnref{eq:H0} and rewrite it in the Lagrangian form
\eq{\label{eq:L[f]}\scL[f]=f^\dagger (\ii\partial_0\sigma^{00}-\ii\partial_1\sigma^{10})f.}
Then according to the Noether theorem, the conserved currents associated with the emergent $\U(1)_\phi\times\U(1)_\theta$ symmetry are given by
\eq{\label{eq:Noether}J_\mu^\phi=\frac{\delta\scL}{\delta(\partial_\mu f)}\frac{\dd f}{\dd \phi}, J_\mu^\theta=\frac{\delta\scL}{\delta(\partial_\mu f)}\frac{\dd f}{\dd \theta},}
where $\dd f/\dd \phi$ and $\dd f/\dd \theta$ are the rate of change of the field $f$ under symmetry transformation. Based on the infinitesimal symmetry transformation in \eqnref{eq:U(1)transform}, we identify
\eq{\label{eq:f rate}\frac{\dd f}{\dd \phi}=-\frac{\ii}{2}\sigma^{03}f,\frac{\dd f}{\dd \theta}=\frac{\ii}{2}\sigma^{13}f.}
From the Lagrangian $\scL[f]$ of the spinon field in \eqnref{eq:L[f]}, we can evaluate the its variations with respect to $\partial_\mu f$,
\eq{\label{eq:variation}\frac{\delta\scL}{\delta(\partial_0 f)}=f^\dagger(\ii\sigma^{00}),\frac{\delta\scL}{\delta(\partial_1 f)}=f^\dagger(-\ii\sigma^{10}).}
Plugging \eqnref{eq:f rate} and \eqnref{eq:variation} into the conserved current formula \eqnref{eq:Noether}, we found
\eq{J_0^\phi=J_1^\theta=\frac{1}{2}f^\dagger \sigma^{03}f, J_1^\phi=J_0^\theta=-\frac{1}{2}f^\dagger \sigma^{13}f.}
Now we look up the operator correspondence table in \tabref{tab:ops}, we can identify the emergent conserved currents with microscopic spin/dimmer operators
\eq{\label{eq:J=ff}J_0^\phi=J_1^\theta\sim S^z_{\pi}=\sum_i(-)^iS_i^z,J_1^\phi=J_0^\theta\sim \Gamma_{\pi}=\sum_i(-)^iS_i^x S_{i+1}^y.}
This is a key result of our derivation.

The operator correspondence can be further validated by matching the symmetry properties on both side. According to \eqnref{eq:n+in}, one can verify that the angles $\phi$ and $\theta$ must transform under discrete symmetries as in \tabref{tab:symm}(d) in order to be consistent with the symmetry property of the $\O(4)$ vector $\vect{n}$. For example,
\eqs{&e^{\ii\phi}=n_1+\ii n_2\overset{\scT}{\to}(-n_1)+(-\ii)(-n_2)=-(n_1+\ii n_2)=e^{\ii(\phi+\pi)},\\
&e^{\ii\theta}=n_3+\ii n_4\overset{T_x}{\to}n_3+\ii(-n_4)=n_3-\ii n_4=e^{-\ii \theta}.}
Based on the symmetry properties, the conserved currents can be expressed in terms of various different kinds of fields,
\eq{J_\mu^\phi\sim\partial_\mu\phi\sim n_1\partial_\mu n_2-n_2\partial_\mu n_1\sim f^\dagger \sigma^{03}f, J_\mu^\theta\sim\partial_\mu\theta\sim n_3\partial_\mu n_4-n_4\partial_\mu n_3\sim f^\dagger \sigma^{13}f.}
Here $\partial_0=\partial_t=\ii\partial_\tau$ denotes the temporal derivative (with respect to either the real time $t$ or the imaginary time $\tau$) and $\partial_1=\partial_x$ denotes the spatial derivative. Using the symmetry properties of either $\phi,\theta$ or $\vect{n}$ or $f$ fields, we can derive how $J_\mu^\phi$ and $J_\mu^\theta$ transforms under all the discrete symmetries. The results are listed in \tabref{tab:symm}(e). One should note that under time-reversal $\scT$, the real time changes sign $\scT:t\to- t$, while the imaginary time does not $\scT:\tau\to\tau$. Nevertheless $\partial_0=\partial_t=\ii\partial_\tau$ changes sign consistently (since $\scT:\ii\to-\ii$). Similarly under reflection $\scP$, the space coordinate should change sign $\scP:x\to-x$, as a result $\scP:\partial_1\to-\partial_1$. Here we show some examples to illustrate the calculation method
\eqs{J_0^\phi\sim\partial_0\phi&\overset{\scT}{\to}(-\partial_0)(\phi+\pi)=-\partial_0\phi\sim-J_0^\phi,\\
J_0^\phi\sim n_1\partial_0 n_2-n_2\partial_0 n_1&\overset{\scT}{\to}(-n_1)(-\partial_0)(-n_2)-(-n_2)(-\partial_0)(-n_1)=-( n_1\partial_0 n_2-n_2\partial_0 n_1)\sim-J_0^\phi,\\
J_0^\phi\sim f^\dagger \sigma^{03}f&\overset{\scT}{\to}f \sigma^{30}\scK\sigma^{03}\scK\sigma^{30}f^\dagger=-f^\dagger(\sigma^{30}(\sigma^{03})^*\sigma^{30})^\intercal f=-f^\dagger\sigma^{03}f\sim-J_0^\phi.}
All three different ways of calculation show the same result that $\scT:J_0^\phi\to-J_0^\phi$. Similar calculations can be performed for other components of the conserved current. In this way, \tabref{tab:symm}(e) can be derived.

As we have determined the symmetry properties of the conserved currents, the last step is to find microscopic lattice operators that also have identical symmetry properties. We found that the $z$-AFM operator $S_{\pi}^z=\sum_i(-)^i S_{i}^z$ and the off-diagonal dimmer operator $\Gamma_{\pi}=\sum_i(-)^i(S_i^x S_{i+1}^y+S_i^y S_{i+1}^x)$ are such operators that matches the symmetry properties. Their symmetry transformations can be derived from the symmetry definition in \tabref{tab:symm}(a). For example,
\eqs{S_\pi^z=\sum_i(-)^i S_i^z\overset{T_x}{\to}&\sum_i(-)^i S_{i+1}^z =\sum_i(-)^{i-1} S_{i}^z =-\sum_i(-)^{i} S_{i}^z=-S_\pi^z\\
\Gamma_\pi=\sum_i(-)^i (S_i^x S_{i+1}^y+S_i^y S_{i+1}^x) \overset{\scP}{\to}&\sum_i(-)^i (S_{-i}^x S_{-i-1}^y+S_{-i}^y S_{-i-1}^x)\\
&=\sum_i(-)^{-i-1} (S_{i+1}^x S_{i}^y+S_{i+1}^y S_{i}^x)\\
&=-\sum_i(-)^i (S_i^x S_{i+1}^y+S_i^y S_{i+1}^x)\sim-\Gamma_\pi.}
Following similar approach, the symmetry transformations of $S_{\pi}^z$ and $\Gamma_\pi$ can be determined as in \tabref{tab:symm}(f). Comparing \tabref{tab:symm}(e) and \tabref{tab:symm}(f), the symmetry property of $S_\pi^z$ is identical to that of $J_0^\phi$ and $J_1^\theta$, and the symmetry property of $\Gamma_\pi$ is identical to that of $J_1^\phi$ and $J_0^\theta$. Therefore the operator correspondence in \eqnref{eq:J=ff} can be established.

\subsection{Abelian Bosonization}\label{sec:bosonization}
Having established the operators corresponding to the emergent conserved currents, we switch gears to calculate the scaling dimension of various other operators at the critical point. We will use Abelian bosonization techniques to analyze the critical point. As proposed in the main text, the critical point can be described by $\O(4)$ NLSM with $k=1$ WZW term deformed by the anisotropy $\mu$ (with $\mu>0$).
\eqs{\label{eq:NLSM1}\scL[\vect{n}]=&\frac{1}{2\kappa}(\partial_\mu\vect{n})^2
+\frac{\ii k}{2\pi^2} \epsilon^{abcd}n_a\partial_\tau n_b\partial_x n_c\partial_u n_d+\mu(n_1^2+n_2^2-n_3^2-n_4^2)\\
&+\lambda (n_3^2-n_4^2)+\lambda' (n_1^2-n_2^2)+\cdots,}
The $\lambda$ and $\lambda'$ terms are also allowed by the microscopic symmetry. $\lambda$ is the driving parameter of the $z$-FM--VBS transition and $\lambda'$ is an irrelevant perturbation at the critical point. Therefore both $\lambda$ and $\lambda'$ are effectively zero at the critical point.

It is not obvious how to treat \eqnref{eq:NLSM1} using the standard bosonization technique. So we first turn to an equivalent fermionic spinon description and then bosonize the fermionic theory. Following the discussion of the fractionalization scheme, the NLSM without anisotropy is dual to an interacting fermionic spinon chain given by \eqnref{eq:Hf}. We can rewrite the Hamiltonian in terms of the left- and right-moving fermion modes $\psi_L=(\psi_{L\uparrow},\psi_{L\downarrow})^\intercal$ and $\psi_R=(\psi_{R\uparrow},\psi_{R\downarrow})^\intercal$,
\eq{H_{\psi}=\int\dd x(\psi_L^\dagger\ii\partial_x\psi_L-\psi_R^\dagger\ii\partial_x\psi_R)+U\vect{J}_L\cdot\vect{J}_R,}
where $\psi_L$ and $\psi_R$ are related to the original spinon field $f$ by
\eq{\label{eq:psif}\mat{\psi_L\\\psi_R}=\frac{1}{\sqrt{2}}\mat{1&-1\\1&1}\mat{f_A\\f_B}.}
The transformation is found by diagonalizing the free spinon Hamiltonian $H_0$ in \eqnref{eq:H0}. The $\SU(2)$ gauge currents $\vect{J}_L$ and $\vect{J}_R$ are simply given by
\eq{\vect{J}_L=(\Re \psi_L^\intercal\ii\sigma^{y}\psi_L,\Im \psi_L^\intercal\ii\sigma^{y}\psi_L,\psi_L^\dagger \psi_L),\quad \vect{J}_R=(\Re \psi_R^\intercal\ii\sigma^{y}\psi_R,\Im \psi_R^\intercal\ii\sigma^{y}\psi_R,\psi_R^\dagger \psi_R).}
Therefore the Hamiltonian can be expanded into
\eq{\label{eq:Hpsi}H_{\psi}=\int\dd x(\psi_L^\dagger\ii\partial_x\psi_L-\psi_R^\dagger\ii\partial_x\psi_R)+\frac{U_\pm}{2}\big((\psi_L^\intercal\ii\sigma^{y}\psi_L)^\dagger(\psi_R^\intercal\ii\sigma^{y}\psi_R)+\text{h.c.}\big)+U_3 \psi_L^\dagger \psi_L\psi_R^\dagger \psi_R,}
where $U_\pm$ and $U_3$ are expected to be the same as $U_\pm=U_3=U$. Using the basis transformation \eqnref{eq:psif}, we can change all the symmetry transforms to the $\psi$ fermion basis, as listed in \tabref{tab:symm}(g). We can further translate the operator correspondence in \tabref{tab:ops} to \tabref{tab:ops2}. In particular, the $\O(4)$ vector $\vect{n}=(n_1,n_2,n_3,n_4)$ couples to the fermionic spinon as
\eq{n_1 S_0^x+n_2 S_\pi^y+n_3 S_0^z+n_4 \Psi_\pi=\psi_L^\dagger(n_1\sigma^{x}+n_2 \sigma^{y}+n_3 \sigma^{z}-\ii n_4)\psi_R+\text{h.c.}.}

\begin{table}[htp]
\caption{Spin and dimmer operators in terms of fermionic spinon field bilinears}
\begin{center}
\begin{tabular}{ccc}
& $Q=0$ & $Q=\pi$\\
\hline
$S^x_Q$ & $\psi_L^\dagger\sigma^x\psi_R+\psi_R^\dagger\sigma^x\psi_L$ & $\psi_L^\dagger\sigma^x\psi_L+\psi_R^\dagger\sigma^x\psi_R$\\
$S^y_Q$ & $\psi_L^\dagger\sigma^y\psi_L+\psi_R^\dagger\sigma^y\psi_R$ & $\psi_L^\dagger\sigma^y\psi_R+\psi_R^\dagger\sigma^y\psi_L$\\
$S^z_Q$ & $\psi_L^\dagger\sigma^z\psi_R+\psi_R^\dagger\sigma^z\psi_L$ & $\psi_L^\dagger\sigma^z\psi_L+\psi_R^\dagger\sigma^z\psi_R$\\
$\Psi_Q$ & $-\psi_L^\dagger\psi_L+\psi_R^\dagger\psi_R$ & $-\ii\psi_L^\dagger\psi_R+\ii\psi_R^\dagger\psi_L$\\
$\Gamma_Q$ & $-\ii\psi_L^\dagger\sigma^z\psi_R+\ii\psi_R^\dagger\sigma^z\psi_L$ & $\psi_L^\dagger\sigma^z\psi_L-\psi_R^\dagger\sigma^z\psi_R$
\end{tabular}
\end{center}
\label{tab:ops2}
\end{table}

Now we can bosonize the fermionic spinon Hamiltonian $H_\psi$ in \eqnref{eq:Hpsi} by defining the boson field $\varphi=(\varphi_{L\uparrow},\varphi_{L\downarrow},\varphi_{R\uparrow},\varphi_{R\downarrow})^\intercal$ via
\eq{\psi_{\alpha\sigma}=\frac{\kappa_{\alpha\sigma}}{\sqrt{2\pi}}e^{\ii\varphi_{\alpha\sigma}},\quad(\alpha=L,R;\sigma=\uparrow,\downarrow)}
where $\kappa_{\alpha\sigma}$ is the Klein factor that ensures the anticommutation of the fermion operators. The density fluctuations are given by
\eq{\psi_{\alpha\sigma}^\dagger\psi_{\alpha\sigma}=\frac{1}{2\pi}(-)^\alpha\partial_x\varphi_{\alpha\sigma},\quad \text{where }(-)^\alpha=\Big\{\begin{array}{ll}+1&\alpha=L,\\-1&\alpha=R.\end{array}}
With this setup, we can show that
\eqs{&\frac{U_\pm}{2}\big((\psi_L^\intercal\ii\sigma^{y}\psi_L)^\dagger(\psi_R^\intercal\ii\sigma^{y}\psi_R)+\text{h.c.}\big)=-\frac{U_\pm}{2\pi^2}(e^{\ii(-\varphi_{L\uparrow}-\varphi_{L\downarrow}+\varphi_{R\uparrow}+\varphi_{R\downarrow})}+\text{h.c.})=-u_\pm \cos(l_0^\intercal \varphi),\\
&U_3 \psi_L^\dagger \psi_L\psi_R^\dagger \psi_R=-\frac{U_3}{4\pi^2}(\partial_x\varphi_{L\uparrow}+\partial_x\varphi_{L\downarrow})(\partial_x\varphi_{R\uparrow}+\partial_x\varphi_{R\downarrow})=-\frac{u_3}{4\pi}\partial_x\varphi^\intercal\smat{0&0&1&1\\0&0&1&1\\1&1&0&0\\1&1&0&0}\partial_x\varphi,}
where $l_0^\intercal=(1,1,-1,-1)$ and $u_\pm=U_\pm/\pi^2$,$u_3=U_3/\pi$. Given that the interaction $U$ is introduce to penalize the gauge current fluctuation, it is expected that $U_\pm=U_3=U>0$ and hence $u_\pm=U_\pm/\pi^2>0$, $u_3=U_3/\pi>0$. Then $H_\psi$ can be bosonized to a Luttinger liquid (LL) theory described by the following Lagrangian density
\eq{\scL[\varphi]=\frac{1}{4\pi}(\partial_\tau\varphi^\intercal K \partial_x\varphi+\partial_x\varphi^\intercal V\partial_x\varphi)-u_\pm \cos(l_0^\intercal \varphi),}
where the $K$ and $V$ matrices are given by
\eq{K=\smat{1&&&\\&1&&\\&&-1&\\&&&-1},\quad V=\smat{1&&-u_3&-u_3\\&1&-u_3&-u_3\\-u_3&-u_3&1&\\-u_3&-u_3&&1}.}
In the case of $u_3>0$, the scaling dimension $\Delta_0$ of $\cos(l_0^\intercal \varphi)$ is given by
\eq{\Delta_0=2\sqrt{\frac{1-2u_3}{1+2u_3}}<2,}
indicating that the $\cos(l_0^\intercal \varphi)$ is relevant. $u_\pm\to+\infty$ under renormalization group (RG) flow given by the following flow equation
\eq{\frac{\dd}{\dd \ell}u_\pm=(2-\Delta_0)u_\pm, \frac{\dd}{\dd\ell}\Delta_0^{-1}=u_\pm^2.}
At the new RG fixed point, the scaling dimension $\Delta_0=0$, from which we can infer $u_3=1/2$ and hence the $V$ matrix becomes
\eq{\label{eq:VSO(4)}V=\smat{1&&-1/2&-1/2\\&1&-1/2&-1/2\\-1/2&-1/2&1&\\-1/2&-1/2&&1}.}
One can check that $l_0^\intercal K^{-1}l_0=0$, which indicates $e^{\ii l_0^\intercal\varphi}$ is a bosonic operator. So as $u_\pm$ flows to infinity, the field $\varphi$ will be pinned by the cosine term to $l_0^\intercal\varphi=0\mod 2\pi$. Any operator $\scO_l=e^{\ii l^\intercal\varphi}$ which does not commute with $\cos(l_0^\intercal \varphi)$ (i.e.~$l^\intercal K^{-1}l_0\neq 0$) will be gapped out. Using this criterion, it is easy to check that all fermions are gapped out. The remaining gapless operators correspond to the $\O(4)$ vector $\vect{n}$. To see this, we can first bosonize the $\vect{n}$ as follows
\eqs{&e^{\ii\phi}=n_1+\ii n_2\sim \psi_{L\uparrow}^\dagger\psi_{R\downarrow}+\psi_{R\uparrow}^\dagger\psi_{L\downarrow}=\frac{1}{2\pi}(e^{\ii l_1^\intercal \varphi}+e^{\ii l_2^\intercal \varphi}),\\
&e^{\ii\theta}=n_3+\ii n_4\sim \psi_{L\uparrow}^\dagger\psi_{R\uparrow}-\psi_{R\downarrow}^\dagger\psi_{L\downarrow}=\frac{1}{2\pi}(e^{\ii l_3^\intercal \varphi}-e^{\ii l_4^\intercal \varphi}),}
with the charge vectors $l_{1,2,3,4}$ given by
\eq{(l_1,l_2,l_3,l_4)=\smat{-1&0&-1&0\\0&1&0&1\\0&-1&1&0\\1&0&0&-1}.}
It is easy to verify that $l_i^\intercal K^{-1}l_0=0$, so the $\vect{n}$ field remains gapless at the RG fixed point. As $l_2=l_1+l_0$ and $l_4=l_3+l_0$, we can establish the following equivalence relations $e^{\ii l_1^\intercal \varphi}\sim e^{\ii l_2^\intercal \varphi}\sim e^{\ii\phi}$ and $e^{\ii l_3^\intercal \varphi}\sim e^{\ii l_4^\intercal \varphi}\sim e^{\ii\theta}$. So at the RG fixed point, there are only two independent bosonic modes $\phi$ and $\theta$. The effective $K$ matrix for these two modes can be obtained from the projection $K_\text{eff}^{-1}=P^\intercal K^{-1} P$ with $P=(l_1,l_3)$. The result is
\eq{K_\text{eff}=\mat{0&1\\1&0}.}
This exactly describes a bosonic CFT with central charge $c=1$, corresponding to the spin $\SU(2)_1$ CFT. The spin $\SU(2)$ structure is not obvious in the Abelian bosonization theory. But the fact the $\O(4)$ vector $\vect{n}$ has the scaling dimension
\eq{\Delta_\vect{n}=\frac{2u_3}{1+2u_3-\sqrt{1-4u_3^2}}=\frac{1}{2}}
matches the property of the spin $\SU(2)_1$ CFT.

Having established the Luttinger liquid description of the isotropic limit of the $\O(4)$ NLSM, we can deform the theory by the anisotropy
\eqs{\mu(n_1^2+n_2^2-n_3^2-n_4^2) =& 2\mu(\psi_{L\uparrow}^\dagger\psi_{L\uparrow}-\psi_{L\downarrow}^\dagger\psi_{L\downarrow})(\psi_{R\uparrow}^\dagger\psi_{R\uparrow}-\psi_{R\downarrow}^\dagger\psi_{R\downarrow})\\
=&-\frac{\mu}{2 \pi^2}(\partial_x\varphi_{L\uparrow}-\partial_x\varphi_{L\downarrow})(\partial_x\varphi_{R\uparrow}-\partial_x\varphi_{R\downarrow}),}
and investigate the scaling dimension of different operators. The anisotropy $\mu$ further dress the $V$ matrix in \eqnref{eq:VSO(4)} to
\eq{V=\mat{1&&-\frac{1}{2}-\frac{u}{2}&-\frac{1}{2}+\frac{u}{2}\\&1&-\frac{1}{2}+\frac{u}{2}&-\frac{1}{2}-\frac{u}{2}\\-\frac{1}{2}-\frac{u}{2}&-\frac{1}{2}+\frac{u}{2}&1&\\-\frac{1}{2}+\frac{u}{2}&-\frac{1}{2}-\frac{u}{2}&&1},}
where $u=8\mu/\pi>0$ (as $\mu>0$ is expected to favor the $z$-FM and VBS ordering). A generic vortex operator takes the form of
\eq{\scO_l=e^{\ii l^\intercal\varphi},}
labeled by a charge vector $l$ whose components are integers. Given the $V$ matrix, the scaling dimension $\Delta_l$ of $\scO_l$ can be calculated as
\eq{\Delta_l=\frac{1}{4\sqrt{1-u^2}}l^\intercal\mat{1 & -1 & u & -u \\ -1 & 1 & -u & u \\ u & -u & 1 & -1 \\ -u & u & -1 & 1 \\}l.}
With this, we can evaluate the scaling dimension of all operators. We first study vertex operators,
\eqs{\label{eq:vopt}\cmat{n_1=S_0^x\\n_2=S_\pi^y}&\sim\cmat{\psi_{L\uparrow}^\dagger\psi_{R\downarrow}\\
\psi_{L\downarrow}^\dagger\psi_{R\uparrow}\\
\psi_{R\uparrow}^\dagger\psi_{L\downarrow}\\
\psi_{R\downarrow}^\dagger\psi_{L\uparrow}}=\cmat{\scO_{(-1,0,0,1)}\\ \scO_{(0,-1,1,0)}\\ \scO_{(0,1,-1,0)}\\ \scO_{(1,0,0,-1)}}\Rightarrow \Delta=\frac{1}{2}\sqrt{\frac{1+u}{1-u}}=\frac{1}{g},\\
\cmat{n_3=S_0^z\\n_4=\Psi_\pi}&\sim\cmat{\psi_{L\uparrow}^\dagger\psi_{R\uparrow}\\
\psi_{L\downarrow}^\dagger\psi_{R\downarrow}\\
\psi_{R\uparrow}^\dagger\psi_{L\uparrow}\\
\psi_{R\downarrow}^\dagger\psi_{L\downarrow}}=\cmat{\scO_{(-1,0,1,0)}\\ \scO_{(0,-1,0,1)}\\ \scO_{(1,0,-1,0)}\\ \scO_{(0,1,0,-1)}}\Rightarrow \Delta=\frac{1}{2}\sqrt{\frac{1-u}{1+u}}=\frac{g}{4},\\
\cmat{S_\pi^x\\S_0^y}&\sim\cmat{\psi_{L\uparrow}^\dagger\psi_{L\downarrow}\\
\psi_{L\downarrow}^\dagger\psi_{L\uparrow}\\
\psi_{R\uparrow}^\dagger\psi_{R\downarrow}\\
\psi_{R\downarrow}^\dagger\psi_{R\uparrow}}=\cmat{\scO_{(-1,1,0,0)}\\ \scO_{(1,-1,0,0)}\\ \scO_{(0,0,-1,1)}\\ \scO_{(0,0,1,-1)}}\Rightarrow \Delta=\frac{1}{\sqrt{1-u^2}}=\frac{1}{g}+\frac{g}{4}.}
Here we have introduced another Luttinger parameter $g=2\sqrt{(1-u)/(1+u)}$ in replacement of $u$. We then investigate current operators,
\eqs{\label{eq:copt}S_\pi^z&\sim\psi_{L\uparrow}^\dagger\psi_{L\uparrow}-\psi_{L\downarrow}^\dagger\psi_{L\downarrow}+\psi_{R\uparrow}^\dagger\psi_{R\uparrow}-\psi_{R\downarrow}^\dagger\psi_{R\downarrow}=\frac{1}{2\pi}\partial_x(\varphi_{L\uparrow}-\varphi_{L\downarrow}-\varphi_{R\uparrow}+\varphi_{R\downarrow})=\frac{1}{2\pi}\partial_x(l_{34}^\intercal\varphi)\Rightarrow \Delta=1,\\
\Gamma_\pi&\sim\psi_{L\uparrow}^\dagger\psi_{L\uparrow}-\psi_{L\downarrow}^\dagger\psi_{L\downarrow}-\psi_{R\uparrow}^\dagger\psi_{R\uparrow}+\psi_{R\downarrow}^\dagger\psi_{R\downarrow}=\frac{1}{2\pi}\partial_x(\varphi_{L\uparrow}-\varphi_{L\downarrow}+\varphi_{R\uparrow}-\varphi_{R\downarrow})=\frac{1}{2\pi}\partial_x(l_{12}^\intercal\varphi)\Rightarrow \Delta=1,\\
\Psi_0&\sim-\psi_{L\uparrow}^\dagger\psi_{L\uparrow}-\psi_{L\downarrow}^\dagger\psi_{L\downarrow}+\psi_{R\uparrow}^\dagger\psi_{R\uparrow}+\psi_{R\downarrow}^\dagger\psi_{R\downarrow}=-\frac{1}{2\pi}\partial_x(\varphi_{L\uparrow}+\varphi_{L\downarrow}+\varphi_{R\uparrow}+\varphi_{R\downarrow})=-\frac{1}{2\pi}\partial_x(l_{D}^\intercal\varphi),}
where $l_{12}=(1,-1,1,-1)^\intercal$, $l_{34}=(1,-1,-1,1)^\intercal$, $l_{D}=(1,1,1,1)^\intercal$. Since $l_{12}^\intercal K^{-1}l_0=l_{34}^\intercal K^{-1}l_0$, the $l_{12}^\intercal\varphi$ and $l_{34}^\intercal\varphi$ fluctuations remain gapless. The scaling dimension of $\partial_x(l_{12}^\intercal\varphi)$ and $\partial_x(l_{34}^\intercal\varphi)$ are both $\Delta=1$, as they correspond to conserved currents. On the other hand, $l_{D}^\intercal K^{-1}l_0\neq0$ indicates that $l_{D}^\intercal\varphi$ is gapped and therefore the operator $\partial_x(l_D^\intercal\varphi)$ does not have a scaling dimension. Finally, we can calculate the scaling dimension of the perturbation $n_1^2-n_2^2$ and $n_3^2-n_4^2$,
\eqs{n_1^2-n_2^2&=-2\psi_{L\uparrow}^\dagger\psi_{L\downarrow}\psi_{R\uparrow}^\dagger\psi_{R\downarrow}+\text{h.c.}=-\frac{1}{\pi^2}\cos(l_{12}^\intercal\varphi)\Rightarrow \Delta=2\sqrt{\frac{1+u}{1-u}}>2,\\
n_3^2-n_4^2&=-2\psi_{L\uparrow}^\dagger\psi_{L\downarrow}\psi_{R\uparrow}\psi_{R\downarrow}^\dagger+\text{h.c.}=-\frac{1}{\pi^2}\cos(l_{34}^\intercal\varphi)\Rightarrow \Delta=2\sqrt{\frac{1-u}{1+u}}<2.}
With $\mu>0$ (and hence $u>0$), $\lambda(n_3^2-n_4^2)$ is relevant and $\lambda'(n_1^2-n_2^2)$ is irrelevant. An quick argument is that at the $\O(4)$ isotropic point ($\lambda=\lambda'=\mu=0$) the field theory is equivalent to the $\SU(2)_1$ conformal field theory (CFT), where $\lambda$ and $\lambda'$ terms are both in the symmetric tensor representation of the $\O(4)$ group which are marginal. Because the $\SU(2)_1$ CFT is described by the Hamiltonian $\scH=\vect{J}_L^2+\vect{J}_R^2$ in terms of the $\SU(2)_L\times \SU(2)_R$ current operators. The symmetric tensor transforms as the $(\mathbf{1}_L,\mathbf{1}_R)$ representation under $\SU(2)_L\times \SU(2)_R$,  corresponding to the $J_L^a J_R^b$ ($a,b=1,2,3$) operators, which share the same scaling dimension as $\scH$ and are therefore marginal (in fact marginally relevant). Away from this limit, the $\mu>0$ anisotropy will enhance the $(n_3, n_4)$ fluctuation and suppress the $(n_1,n_2)$ fluctuation, making $\lambda$ relevant and $\lambda'$ irrelevant. Therefore $\lambda$ is identified as the driving parameter of the DQCP. As the driving parameter $\lambda$ couples to $n_3^2-n_4^2$, so $\lambda$ must scale with the correlation length $\xi$ as
\eq{\lambda\sim\xi^{-\big(2-2\sqrt{\frac{1-u}{1+u}}\big)}\Rightarrow\xi\sim\lambda^{-\frac{1}{2\big(1-\sqrt{\frac{1-u}{1+u}}\big)}}=\lambda^{-\nu}.}
Therefore the critical exponent $\nu$ is given by
\eq{\nu=\frac{1}{2\Big(1-\sqrt{\frac{1-u}{1+u}}\Big)}=\frac{1}{2-g}.}
The scaling dimensions in \eqnref{eq:vopt} and \eqnref{eq:copt} can be interpreted as the correlation function. Suppose $A$ is a spin/dimmer operator ($A=S^x,S^y,S^z,\Psi,\Gamma$), its correlation function is generally given by
\eq{G_A(r)=\langle A_i A_{i+r}\rangle\sim\frac{c_1}{r^{2\Delta[A_0]}}+\frac{c_2(-)^r}{r^{2\Delta[A_\pi]}},}
where $c_{1,2}$ are constants (which will be omitted in the following) and $\Delta[A]$ denotes the scaling dimension of the operator $A$. Using this formula, the results in \eqnref{eq:vopt} and \eqnref{eq:copt} imply the following correlation functions
\eqs{
G_x(r)=\langle S_i^xS_{i+r}^x\rangle&\sim\frac{1}{r^{2/g}}+\frac{(-)^r}{r^{2/g+g/2}},\\
G_y(r)=\langle S_i^yS_{i+r}^y\rangle&\sim\frac{1}{r^{2/g+g/2}}+\frac{(-)^r}{r^{2/g}},\\
G_z(r)=\langle S_i^zS_{i+r}^z\rangle&\sim\frac{1}{r^{g/2}}+\frac{(-)^r}{r^2},\\
G_\Psi(r)=\langle \Psi_{i}\Psi_{i+r}\rangle&\sim\frac{(-)^r}{r^{g/2}},\\
G_\Gamma(r)=\langle \Gamma_{i}\Gamma_{i+r}\rangle&\sim\frac{(-)^r}{r^2}.}

\end{document}